\documentclass[aps,prd,amsmath,amsfonts,amssymb,eqsecnum,nofootinbib,
  floatfix,secnumarabic,preprintnumbers,nobalancelastpage,
 onecolumn,notitlepage]{revtex4-1}
\pdfoutput=1
\usepackage[utf8x]{inputenc}

\usepackage{subfigure,color,graphicx,hyperref,dsfont,slashed,multirow}
\usepackage{color}
\usepackage{hyperref}
\usepackage{fontenc}
\usepackage{pbox}
\usepackage{feynmp}

\usepackage{mathtools,nccmath}%
\usepackage{ etoolbox, xparse} 

\usepackage{array}
\newcolumntype{P}[1]{>{\centering\arraybackslash}p{#1}}

\DeclarePairedDelimiterX{\abs}[1]\lvert\rvert{\ifblank{#1}{\,\cdot\,}{#1}}

\let\oldabs\abs
\def\abs{\futurelet\testchar\MaybeOptArgAbs}
\def\MaybeOptArgAbs{\ifx[\testchar\let\next\OptArgAbs
	\else \let\next\NoOptArgAbs\fi \next}
\def\OptArgAbs[#1]#2{\oldabs[#1]{#2}}
\def\NoOptArgAbs#1{\ifblank{#1}{\oldabs{}}{\oldabs[\big]{#1}}}

\DeclarePairedDelimiterX{\set}[1]\{\}{\setargs{#1}}
\NewDocumentCommand{\setargs}{>{\SplitArgument{1}{;}}m}
{\setargsaux#1}
\NewDocumentCommand{\setargsaux}{mm}
{\IfNoValueTF{#2}{#1}{\nonscript\,#1\nonscript\;\delimsize\vert\nonscript\:\allowbreak #2\nonscript\,}}
\let\oldset\set
\def\set{\futurelet\testchar\MaybeOptArgSet}
\def\MaybeOptArgSet{\ifx[\testchar \let\next\OptArgSet
	\else \let\next\NoOptArgSet \fi \next}
\def\OptArgSet[#1]#2{\oldset[#1]{#2}}
\def\NoOptArgSet#1{\OptArgSet[\big]{#1}}

\hypersetup{bookmarks=true,unicode=true,pdftoolbar=true,pdfmenubar=true,
  pdffitwindow=false,pdfstartview={FitH},pdftitle={mBLRISS},
  pdfauthor={~M.~Frank,\"{O}.~\"{O}zdal}, pdfsubject={Subject},
  pdfcreator={Creator},pdfproducer={Producer}, pdfkeywords={Non-minimal 
  supersymmetry}{Beyond the Standard Model Physics}{sneutrino dark matter},
  pdfnewwindow=true,colorlinks=true,
  linkcolor=blue,citecolor=magenta,filecolor=magenta,urlcolor=cyan}
\newcommand{\be}{\begin{equation}}
\newcommand{\ee}{\end{equation}}
\def\bsp#1\esp{\begin{split}#1\end{split}}
\renewcommand{\figureautorefname}{Fig.}

\def\sectionautorefname~#1\null{Sec.~(#1)\null}
\def\subsectionautorefname~#1\null{sub--Sec.~(#1)\null}
\def\figureautorefname~#1\null{Fig.~#1\null}
\def\tableautorefname~#1\null{Table~#1\null}
\def\equationautorefname~#1\null{Eq.~#1\null}

\begin{document}

\preprint{CUMQ/HEP 198}

{\title{ Relaxing LHC constraints on the $W_R$ mass}

\author{Mariana Frank$^1$\footnote{Email: mariana.frank@concordia.ca}}
\author{\"{O}zer \"{O}zdal$^1$\footnote{Email: ozer.ozdal@concordia.ca}}
\author{Poulose Poulose$^2$\footnote{Email: poulose@iitg.ac.in}}
\affiliation{ $^1$Department of Physics,  
Concordia University, 7141 Sherbrooke St. West ,
Montreal, Quebec, Canada H4B 1R6,}
\affiliation{ $^2$Department of Physics, Indian Institute of Technology Guwahati, Assam 781039, India}

\date{\today}
%\vspace{10pt}
\begin{abstract}
 {We study mass bounds of the $W_R$ gauge boson in generic  left-right symmetric models. Assuming that the gauge bosons couple universally to quarks and leptons, we allow different gauge couplings $g_R \ne g_L$ and mass mixing, $V_{CKM}^L \ne V_{CKM}^R$ in the left and right sectors.  Imposing constraints from collider experiments and $K^0$, $B_d$, $B_s$  physics, we investigate scenarios where $W_R$ is lighter, or heavier than the right handed neutrino  $\nu_R$. In these scenarios,   $W_R$ mass bounds can be considerably relaxed, while $Z_R$ mass bounds are much more stringent. In the case where $M_{W_R} \le M_{\nu_R}$, the experimental constraints come from $W_R \to tb $ and $W_R \to  jj$ channels, while if  $M_{W_R} \ge M_{\nu_R}$, the dominant constraints come from $W_R \to \ell \ell jj $. The observed (expected) limits in the two-dimensional ($M_{W_R}$, $M_{\nu_R}$)  mass plane excluded at 95\% confidence level extend to approximately $M_{W_R}$= 3.1 (3.3) TeV in the $ee$ channel and 3.3 (3.4) TeV in the ($\mu\mu$) channel, for a large range of right-handed neutrino masses up to $M_{\nu_R}$= 2.1 (2.1) TeV in the $ee$ channel and 2.6 (2.5) in the  ($\mu\mu$) channel, representing a significant relaxation of the mass bounds.}
\end{abstract} 

%Use showkeys class option if keyword

\maketitle
%\flushbottom

%%%%%%%%%%%%%%%%%%%%%%%%%%%%%%%%%%%%%%%%%%%%%%%%%%%%%%%%%%%%%%%%%%%%%%%%%%%%%%
\section{Introduction}\label{sec:intro}

The discovery of the  Higgs boson at the LHC, while providing the missing ingredient  of the Standard Model (SM), has intensified the search for physics beyond it. Though there is a marked lack of non-standard signals at colliders, the existence of neutrino masses is a definite sign of beyond the SM. Using only left-handed neutrinos and the standard Higgs doublet representation of the SM, massive neutrinos appear only from higher dimension operators acting at the Planck scale. While by no means the only solution, left-right symmetric models (LRSMs) \cite{Pati:1974yy,Mohapatra:1974gc,Mohapatra:1974hk}  provide a natural explanation for neutrino masses without resorting to higher scales. Based on the $SU(2)_L \otimes SU(2)_R \otimes U(1)_{B-L}$, these models restore parity symmetry, which is conserved at high energy, and spontaneously broken at some energy scale connected to the $SU(2)_R$ breaking scale \cite{Senjanovic:1975rk}. That is,  LRSMs provide a renormalizable framework where left- and right-handed fields are treated the same way: they are doublets under $SU(2)$. Small neutrino masses are induced by heavy (and most often, Majorana) right-handed neutrinos through the phenomena known as the seesaw mechanism \cite{Mohapatra:1979ia,Mohapatra:1980yp}. Within the framework described here both Type I and Type II seesaw mechanisms can be naturally imbedded in the model. 

In addition, LRSMs gauge the anomaly-free $B-L$ ($B$ - baryon, $L$ - lepton, number) symmetry, and replace the physically meaningless hypercharge $Y$ by the $B-L$ quantum number in the definition of charge
$$Q=T_{3L}+T_{3R}+\frac{B-L}{2},$$
with $T_3$ the third component of isospin.  In some LRSMs,  CP-violating phases are connected with the right-handed quark system, and the smallness of the CP violation in the quark sector is related to the suppression of the $V+A$ currents and the large right-handed gauge boson mass \cite{Mohapatra:1986uf,Mohapatra:1977rj}. These models  also provide a solution to the strong CP violation problem. In LRSMs, the strong CP parameter $\Theta \to - \Theta$ under parity, and thus $\Theta =0$ at tree level, with small non-zero contributions arising at two loop level only \cite{Beg:1978mt}. The LRSMs are constrained at low energies by flavor-violating mixing and decays, by CP observables, and, more recently,  by increasingly restrictive collider limits on new particle (additional gauge or Higgs bosons) masses.

 In making any specific predictions, there are several sources of uncertainty in the model. One comes from the precise nature of the additional Higgs bosons needed to break the symmetry to the SM one.  Higgs triplets are favored by the seesaw mechanisms, and yield Majorana neutrinos. Alternatively, additional  left- and right- handed doublet Higgs representations yield Dirac neutrino masses. Another source of uncertainty comes from the right-handed quark mixing matrix, similar  to that of the left-handed quark Cabibbo-Kobayashi-Maskawa (CKM) mixing.  Two scenarios of left-right models have been most often considered: manifest and pseudo-manifest left-right symmetry.
 Manifest left-right symmetry, assumes no spontaneous CP violation, {\it i.e.} all Higgs vacuum expectation values (VEVs) are real. The  quark mass matrices are Hermitian, and the left- and right-handed quark mixings become identical, up to a sign uncertainty of the elements from negative quark masses \cite{Beg:1977ti,Senjanovic:1978ev}. Pseudo-manifest left-right symmetry  assumes that the CP violation comes entirely from spontaneous symmetry breaking of the vacuum and all Yukawa couplings are real \cite{Mohapatra:1977mj,Branco:1982wp}. The quark mass matrices are complex and symmetric, and the right-handed quark mixing is related to the complex conjugate of the CKM matrix multiplied by additional CP phases. However,  studies of the left-right symmetric model with more general structures, where no {\it a priory} assumptions on masses and mixing in the right sector are made,  also exist in the literature \cite{Langacker:1989xa,Barenboim:1996nd,Frank:2010cj}. 
 
 Another source of uncertainty comes from the gauge coupling constant in the right-handed sector. It is commonly assumed that left-right symmetry imposes $g_L=g_R$. If breaking of $SU(2)_R \otimes U(1)_{B-L}$ occurs at a high scale,  at that scale $g_L=g_R$, but below that, the couplings $g_L$ and $g_R$ could evolve differently, and would be different at low energy scales.
  
 Bounds on extra particle masses depend strongly on the above assumptions, in particular on  the size of the right-handed gauge coupling and/or the right-handed CKM matrix elements, and none more so than the charged gauge boson $W_R$. This boson is interesting for several reasons. First, regardless of the other details of the spectrum of left-right models, discovery of a charged gauge boson will be an indication of the presence of an additional $SU(2)$ symmetry group, and if testing its decay products would indicate that it is right-handed, this will be an unambiguous signal for left-right symmetry\footnote{Unlike the discovery of a $Z^\prime$ which can indicate the presence of any variety of additional $U(1)^\prime$ groups.}. Second, the LHC has significantly improved searches for $W_R$ bosons, and the limits on their masses are becoming more stringent. A previous work considered implications for the single $W_R$ production, $pp \to W_R t$ at the LHC, showing that significant signals above the background are possible, if we relax equality between gauge couplings and of CKM matrices in left and right sectors \cite{Frank:2010cj}.
 
Extensive analyses performed indicate problems with both manifest and pseudo-manifest scenarios \cite{Zhang:2007da}. In the manifest LRSM, there are more minimization conditions than the number of VEVs, thus the model suffers from a fine-tuning problem. In  the pseudo-manifest LRSM,  the decoupling limit yields either a model with light left-handed triplet Higgs, (already excluded by the $\rho$ parameter), or a two Higgs doublet model (excluded  due to large tree level flavor-changing neutral currents). It seems that a better way to proceed would be to keep the parameter space of the model as general as possible,  but impose all constraints from collider searches and from flavor and CP bounds from low-energy phenomenology.
 
 In light of recent experimental measurements and interpretations restricting $W_R$ and $Z_R$ boson masses,  we re-consider the scenarios assumed in the analyses, and recast the results using an unrestricted LRSM. For example, considering the CMS analysis, the effect of $g_R < g_L$ can reduce the production cross section, thus bringing down the lower limit obtained on $M_{W_R}$.  This analysis is timely, as more precise analyses are expected to emerge from the Run II at the LHC in the next year or so.
 
We  first discuss the relevant parameters of the model and their relations to the masses and couplings of the states involved, and then  look for suitable parameter choices that can make an impact on the search channel, at the same time ensuring that all constraints, including those from the flavour sector are satisfied. We also demonstrate that in suitably chosen parameter space regions, it is possible to have other decay channels  open up, resulting in a reduction of the branching ratio to the collider signal channel.

Our work is organized as follows: In Section \ref{sec:model}, we briefly describe the left-right symmetric model with an emphasis on the gauge sector. After we summarize the experimental mass bounds on $W_R$, $\nu_R$ and doubly charged Higgs bosons in Section \ref{sec:experiment}, we explain the model implementation and the experimental constraints employed in our analysis in Section \ref{sec:scan}. We analyze two separate cases, $M_{W_R} <M_{\nu_R}$, $M_{W_R}>M_{\nu_R}$, and then present our results on $W_R-\nu_R$ mass exclusion limits in Section \ref{sec:analysis}. Finally we summarize and conclude in Section \ref{sec:conclusion}.
%%%%%%%%%%%%%%%%%%%%%%%%%%%%%%%%%%%%%%%%%%%%%%%%%%%%%%%%%%%%%%%%%%%%%%%%%%%%%%
\section{The Left-Right Symmetric Model}
\label{sec:model}
%%%%%%%%%%%%%%%%%%%%%%%%%%%%%%%%%%%%%%%%%%%%%%%%%%%%%%%%%%%%%%%%%%%%%%%%%%%%%%
    The framework of the model is based on the gauge group $SU(3)_c \otimes SU(2)_L \otimes SU(2)_R \otimes U(1)_{B-L}$. We extend the Standard Model gauge symmetry, first by introducing the $SU(2)_R$ symmetry, then by gauging the $B-L$ quantum number. Both left- and right-handed fermions are  doublets under
the extended $SU(3)_c \otimes SU(2)_L \otimes SU(2)_R \otimes U(1)_{B-L}$ gauge group.  
The quantum numbers of the particles under
 $SU(2)_L \otimes SU(2)_R \otimes U(1)_{B-L}$ are:
 
  \begin{eqnarray}
 L_{Li} = \begin{pmatrix}\nu_L \\ \ell_L\end{pmatrix}_i \, \sim
(\mathbf{2},\mathbf{1},\mathbf{-1}) \, , ~&
L_{Ri} =
\begin{pmatrix}\nu_R \\ \ell_R\end{pmatrix}_i \sim
(\mathbf{1},\mathbf{2},\mathbf{-1}) \, , \qquad {\rm for~the~leptons,~and}
\end{eqnarray}
\begin{eqnarray} 
 Q_{Li} = \begin{pmatrix}u_L \\ d_L\end{pmatrix}_i \, \sim
(\mathbf{2},\mathbf{1},\mathbf{1/3}) \, , ~&
Q_{Ri} =
\begin{pmatrix}u_R \\ d_R\end{pmatrix}_i \sim
(\mathbf{1},\mathbf{2},\mathbf{1/3}) \, ,  \qquad {\rm for~the~quarks,}
\end{eqnarray}
where $i=1,2,3$ are generation indices.  The subscripts $L$ and $R$ are associated with the
projection operators $P_{L,R} = \frac 12 (1 \mp \gamma_5)$.  
The electroweak symmetry is
broken by the bi-doublet Higgs field
\begin{equation}
 \Phi \equiv \begin{pmatrix} \phi_1^0 & \phi_2^+ \\ \phi_1^- & \phi_2^0 \end{pmatrix} \sim (\mathbf{2},\mathbf{2},\mathbf{0})\, .
\end{equation}
In addition, to break the $SU(2)_R \otimes U(1)_{B-L}$ gauge symmetry and to provide Majorana
mass terms for neutrinos (the right-handed neutrino is automatically included in the right-handed lepton doublet), we introduce the Higgs triplets 
\begin{equation}
 \Delta_{L} \equiv \begin{pmatrix} \delta_{L}^+/\sqrt{2} & \delta_{L}^{++} \\ \delta_{L}^0 & -\delta_{L}^+/\sqrt{2} \end{pmatrix} \sim (\mathbf{3},\mathbf{1},\mathbf{2}) \, , ~~~ \Delta_{R} \equiv \begin{pmatrix} \delta_{R}^+/\sqrt{2} & \delta_{R}^{++} \\ \delta_{R}^0 & -\delta_{R}^+/\sqrt{2} \end{pmatrix} \sim (\mathbf{1},\mathbf{3},\mathbf{2})\, .
\end{equation}
While only $\Delta_R$ is needed for symmetry breaking,  $\Delta_L$ is included to preserve left-right symmetry. After symmetry breaking, the most
general vacuum is 
\begin{gather} 
 \langle \Phi\rangle = \begin{pmatrix} \kappa_1/\sqrt{2} & 0 \\ 0 &
\kappa_2e^{i\alpha}/\sqrt{2} \end{pmatrix}, \quad \langle \Delta_{L}
\rangle = \begin{pmatrix} 0 & 0 \\ v_{L}e^{i\theta_L}/\sqrt{2} & 0
\end{pmatrix}, \quad \langle \Delta_{R} \rangle = \begin{pmatrix} 0 &
0 \\ v_{R}/\sqrt{2} & 0 \end{pmatrix}. 
\end{gather} 
We define the ratio $\displaystyle \tan \beta=\frac{\kappa_1}{\kappa_2}$. 
 The Lagrangian density for this model contains  kinetic, Yukawa terms  and potential terms:
\begin{eqnarray}
\mathcal{L}_{\rm{LRSM}}= \mathcal{L}_{\rm{\rm kin}}+\mathcal{L}_{Y}-V(\Phi,\Delta_L, \Delta_R) \,. 
\label{eq:lagrangian}
\end{eqnarray}
The kinetic term is
\begin{eqnarray}
L_{\rm kin}&=&i\sum\bar{\psi}\gamma^\mu D_\mu\psi 
=\bar{L}_L\gamma^{\mu}\left(i\partial_{\mu}+g_{L}\frac{\vec{\tau}}{2}\cdot\vec{W}_{L\mu}-\frac{g_{B-L}}{2}B_{\mu}\right)L_L +\bar{L}_R\gamma^{\mu}\left(i\partial_{\mu}+ g_{R}\frac{\vec{\tau}}{2}\cdot\vec{W}_{R\mu}-\frac{g_{B-L}}{2}B_{\mu}\right)L_R
 \nonumber \\
&+& \bar{Q}_L\gamma^{\mu}\left(i\partial_{\mu}+g_{L}\frac{\vec{\tau}}{2}\cdot\vec{W}_{L\mu}+\frac{g_{B-L}}{6}B_{\mu}\right) Q_L %\nonumber \\
+\bar{Q}_R\gamma^{\mu}\left(i\partial_{\mu}+g_{R}\frac{\vec{\tau}}{2}\cdot \vec{W}_{R\mu}+\frac{g_{B-L}}{6}B_{\mu}\right)Q_R \,,
\label{eq:kinetic}
\end{eqnarray}
where we introduce the gauge fields, $\vec{W}_{L,R}$ and $B$ corresponding to $SU(2)_{L,R}$ and $U(1)_{B-L}$. 
The next term in the \autoref{eq:lagrangian} is the Yukawa interaction for quarks and leptons 
\begin{eqnarray}
\mathcal{L}_Y&=&-\Big[Y_{L_L} {\bar L}_{L} \Phi L_{R} +{\tilde Y}_{L_R} {\bar L}_R \Phi L_L+
Y_{Q_L} {\bar Q}_L {\tilde \Phi} Q_R
  +{\tilde Y}_{Q_R} {\bar Q}_R {\tilde \Phi} Q_L
  +h_{L}^{ij}\overline {L}^c_{L_i} i \tau_2 \Delta_L L_{L_j}\nonumber \\
  &+&h_{R}^{ij}\overline {L}^c_{R_i}i \tau_2 \Delta_R L_{R_j} +\rm{h.c.} \Big]\, , 
\label{fermion_yukawa}
\end{eqnarray}
where   for leptons $Y_{L_{L,R}}$, ${\tilde Y}_{L_{L,R}}$  and for quarks $Y_{Q_{L,R}}$, ${\tilde Y}_{Q_{L,R}}$ are 3$\times 3$ diagonal complex matrices, $h_{L}^{ij}$ and $h_{R}^{ij}$ are $3\times 3$ complex symmetric Yukawa matrices and $\tilde \Phi=\tau_2 \Phi^{\star} \tau_2$.  
Finally, the last term in the Lagrangian in \autoref{eq:lagrangian} is the scalar potential for the bidoublet $\Phi$  and triplet $\Delta_{L,R}$ Higgs fields, and is given by
\begin{eqnarray}
\!\!\! \! V(\phi,\Delta_L,\Delta_R)& = &-\mu_{1}^2\left({\rm Tr}\left[\Phi^\dagger\Phi\right]\right)-\mu_{2}^2\left({\rm Tr}\left[\tilde{\Phi}\Phi^\dagger\right]+\left({\rm Tr}\left[\tilde{\Phi}^\dagger\Phi\right]\right)\right)-\mu_{3}^2\left({\rm Tr}\left[\Delta_L\Delta_L^{\dagger}\right]+{\rm Tr}\left[\Delta_R\Delta_R^{\dagger}\right]\right) \nonumber \\
&+&\lambda_1\left(\left({\rm Tr}\left[\Phi\Phi^\dagger\right]\right)^2\right)
+\lambda_2\left(\left({\rm Tr}\left[\tilde{\Phi}\Phi^\dagger\right]\right)^2
+\left({\rm Tr}\left[\tilde{\Phi}^\dagger\Phi\right]\right)^2\right)
+\lambda_3\left({\rm Tr}\left[\tilde{\Phi}\Phi^\dagger\right]{\rm Tr}\left[\tilde{\Phi}^\dagger\Phi\right]\right)\nonumber\\
&+&\lambda_4\left({\rm Tr}\left[\Phi\Phi^{\dagger}\right]\left({\rm Tr}\left[\tilde{\Phi}\Phi^\dagger\right]
+{\rm Tr}\left[\tilde{\Phi}^\dagger\Phi\right]\right)\right) %\nonumber \\
+\rho_1\left(\left({\rm Tr}\left[\Delta_L\Delta_L^{\dagger}\right]\right)^2
+\left({\rm Tr}\left[\Delta_R\Delta_R^{\dagger}\right]\right)^2\right)\nonumber\\
&+&\rho_2\left({\rm Tr}\left[\Delta_L\Delta_L\right]{\rm Tr}\left[\Delta_L^{\dagger}\Delta_L^{\dagger}\right]
+{\rm Tr}\left[\Delta_R\Delta_R\right]{\rm Tr}\left[\Delta_R^{\dagger}\Delta_R^{\dagger}\right]\right) +
\rho_3\left({\rm Tr}\left[\Delta_L\Delta_L^{\dagger}\right]{\rm Tr}\left[\Delta_R\Delta_R^{\dagger}\right]
\right)\nonumber \\
&+&\rho_4\left({\rm Tr}\left[\Delta_L\Delta_L\right]{\rm Tr}\left[\Delta_R^{\dagger}\Delta_R^{\dagger}\right]
+{\rm Tr}\left[\Delta_L^{\dagger} \Delta_L^{\dagger}\right]{\rm Tr}\left[\Delta_R\Delta_R\right]\right)
%\nonumber\\
+ \alpha_1{\rm Tr}\left[\Phi\Phi^{\dagger}\right]\left({\rm Tr}\left[\Delta_L\Delta_L^{\dagger}\right]
+{\rm Tr}\left[\Delta_R\Delta_R^{\dagger}\right]\right) \nonumber \\
&+&\alpha_2\left({\rm Tr}\left[\Phi\tilde{\Phi}^{\dagger}\right]{\rm Tr}\left[\Delta_R\Delta_R^{\dagger}\right]
+{\rm Tr}\left[\Phi^{\dagger}\tilde{\Phi}\right]{\rm Tr}\left[\Delta_L\Delta_L^{\dagger}\right]\right)
%\nonumber \\
+\alpha_2^{*}\left({\rm Tr}\left[\Phi^{\dagger}\tilde{\Phi}\right]{\rm Tr}\left[\Delta_R\Delta_R^{\dagger}
\right]+{\rm Tr}\left[\tilde{\Phi}^{\dagger}\Phi\right]{\rm Tr}\left[\Delta_L\Delta_L^{\dagger}\right]\right) \nonumber \\
&+& \alpha_3\left({\rm Tr}\left[\Phi\Phi^{\dagger}\Delta_L\Delta_L^{\dagger}\right]
+{\rm Tr}\left[\Phi^{\dagger}\Phi\Delta_R\Delta_R^{\dagger}\right]\right)%\nonumber\\
+\beta_1\left({\rm Tr}\left[\Phi\Delta_R\Phi^{\dagger}\Delta_L^{\dagger}\right]
+{\rm Tr}\left[\Phi^{\dagger}\Delta_L\Phi\Delta_R^{\dagger}\right]\right)\nonumber\\
&+&\beta_2\left({\rm Tr}\left[\tilde{\Phi}\Delta_R\Phi^{\dagger}\Delta_L^{\dagger}\right]
+{\rm Tr}\left[\tilde{\Phi}^{\dagger}\Delta_L\Phi\Delta_R^{\dagger}\right]\right)%\nonumber\\
+\beta_3\left({\rm Tr}\left[\Phi\Delta_R\tilde{\Phi}^{\dagger}\Delta_L^{\dagger}\right]
+{\rm Tr}\left[\Phi^{\dagger}\Delta_L\tilde{\Phi}\Delta_R^{\dagger}\right]\right),
~~~~ \label{eq:pot_htm}
\end{eqnarray}
where we include the complex parameters explicitly.

 The $SU(2)_R$, $SU(2)_L$ and $U(1)_{B-L}$ gauge couplings in \autoref{eq:kinetic} are denoted by $g_R$, $g_L$ and $g_{B-L}$.
 For the LRSM model to break down to the SM,   the hierarchy of the vacuum expectation values (VEVs) must be,  $v_R \gg (\kappa_1,~\kappa_2)\gg v_L$, and $\sqrt{\kappa_1^2+\kappa_2^2} = v = 246$ GeV. Here the  presence of non-zero VEV of $\Delta_R$, $v_R$ breaks the $SU(2)_R\otimes U(1)_{B-L}$ to $U(1)_Y$, while the presence of bi-doublet VEVs $\kappa_1$ and $\kappa_2$ break the electroweak symmetry, at the same time inducing a VEV for $\Delta_L$ denoted by $v_L$.
 The third component of the $SU(2)_R$ gauge boson and the gauge boson corresponding to $U(1)_{B-L}$ mix to give the mass eigenstate $Z_R$ and the gauge boson corresponding to the $U(1)_{Y}$, $B^\mu$. Denoting the mixing angle as $\phi$, 
 \begin{equation}
 \left(\begin{array}{c}Z_R^\mu\\B^\mu\end{array} \right) =
 \left(\begin{array}{cc}\cos\phi&-\sin\phi\\\sin\phi&\cos\phi\end{array}\right) \left(\begin{array}{c}W_R^{3\mu}\\V^\mu\end{array} \right)
 \end{equation}
 Denoting the weak mixing angle as $\theta_W$,  the three neutral gauge bosons further mix to give the mass eigenstates $Z_L^\mu$, $Z_R^\mu$ and the  photon, $A^\mu$:
  \begin{equation}
 \left(\begin{array}{c}Z_L^\mu\\B^\mu\\Z_R^\mu\end{array} \right) =
 \left(\begin{array}{ccc}\cos\theta_W&-\sin\theta_W \sin\phi&-\sin\theta_W\cos\phi\\
 \sin\theta_W&\cos\theta_W\sin\phi&\cos\theta_W\cos\phi\\
 0&\cos\phi&-\sin\phi\end{array}\right) \left(\begin{array}{c}W_L^{3\mu}\\W_R^{3\mu}\\V^\mu\end{array} \right)
 \end{equation}
 yielding 
 \begin{eqnarray}
 M_{A}&=&0 \nonumber \\
 M^2_{Z_{1,2}}&=&\frac14 \Big[ \left[g_L^2 v^2+2v_R^2 (g_R^2+g_{B-L}^2)\right] \mp \sqrt{\left[g_L^2 v^2+2v_R^2 (g_R^2+g_{B-L}^2)\right]^2-4g_L^2(g_R^2+2g_{B-L}^2)v^2v_R^2} \Big] \,.
 \end{eqnarray}
 In the charged sector, the left and right gauge bosons mix to give the mass eigenstates, $W_1$ and $W_2$;
  \begin{equation}
 \left(\begin{array}{c}W_1\\W_2\end{array} \right) =
 \left(\begin{array}{cc}\cos\xi&-\sin\xi\\\sin\xi&\cos\xi\end{array}\right) \left(\begin{array}{c}W_L\\W_R\end{array} \right) \, ,
 \end{equation}
 where the mixing angle $\xi$ and the mass eigenvalues are given by
 \begin{eqnarray}
 \tan 2\xi&=&\frac{4g_Rg_L\kappa_1\kappa_2}{2g_R^2v_R^2+(g_R^2-g_L^2)v^2}\approx \frac{2g_L\kappa_1\kappa_2}{g_Rv_R^2}  \label{tan2xi} \\
 M_{W_1}^2&=&\frac{1}{4}\left[g_L^2v^2\cos^2\xi+g_R^2(2v_R^2+v^2)\sin^2\xi-2g_Rg_L\kappa_1\kappa_2\cos\xi\sin\xi\right] \nonumber \\
 M_{W_2}^2&=&\frac{1}{4}\left[g_L^2v^2\sin^2\xi+g_R^2(2v_R^2+v^2)\cos^2\xi+2g_Rg_L\kappa_1\kappa_2\cos\xi\sin\xi\right]\, .
 \end{eqnarray}
  In the limit of $(\kappa_1, \kappa_2)\ll v_R$ and $g_R\sim g_L$ we have 
 $\displaystyle \sin\xi \approx \frac{\kappa_1\kappa_2}{v_R^2},~\sin^2\xi\approx 0,~~\cos\xi\approx 1 $, leading to 
  \begin{eqnarray}
M^2_{W_1}= \frac14 g_L^2v^2\, , \qquad M_{W_2}^2=\frac{1}{4}\left[2g_R^2v_R^2+g_R^2v^2+2g_Rg_L
\frac{\kappa^2_1\kappa^2_2}{v_R^2} \right].
 \end{eqnarray}
The $SU(2)_R$ breaking scale $v_R$, and mixing angle $\xi$ are restricted from low energy observables, such as 
$K_L-K_S$, $\epsilon_K$, $B^0-\bar{B}^0$ mixings and $b \to s\gamma$ processes, which constrain the right scale 
through the charged right handed $W_R$ boson mass as well as the triplet Higgs masses. Taking into account the smallness of the mixing angle $\xi$, in what follows we shall refer to $W_1$ as $W_L$ and $W_2$ as $W_R$, as to be able to compare with experimental results and nomenclature\footnote{Note however that they both contain a non-zero left and right  $SU(2)$ gauge component.}.
%\clearpage
%%%%%%%%%%%%%%%%%%%%%%%%%%%%%%%%%%%%%%%%%%%%%%%%%%%%%%%%%%%%%%%%%%%%%%%%%%%%%
\section{Mass Bounds on right-handed gauge, Higgs bosons, and neutrinos from colliders}
\label{sec:experiment}
The choice of the parameter space of the model would affect masses in the right-handed sector, and thus below,  we summarize collider limits imposed on these.
%%%%%%%%%%%%%%%%%%%%%%%%%%
\subsection{Right-handed charged gauge boson}
\label{subsec:WR}
%%%%%%%%%%%%%%%%%%%%%%%%
The $W_{R}$ boson can decay into jets and, if the $SU(2)_{\rm R}$
gauge coupling equals the $SU(2)_{L}$ one ($g_L=g_R$), limits on sequential
$W^{\prime}$ bosons can be reinterpreted straightforwardly. ATLAS and CMS have
obtained bounds of $3.6$~TeV~\cite{Aaboud:2017yvp} and $3.3$~TeV~\cite{Sirunyan:2018xlo}, on such a sequential extra gauge boson, respectively. 
%Recent results at the LHC \cite{Sirunyan:2017ukk,201839} have put bounds on $pp \to t {\bar b}$, assuming this to be the dominant decay of $W_R$.
LHC has analyzed possible signatures of the additional gauge bosons in LRSM, which for the charged gauge boson  translate into  three main search channels, (i) $W_R\rightarrow t \bar b$ \cite{Sirunyan:2017ukk, 201839}, (ii) $W_R\rightarrow 2j$ \cite{Aaboud:2017yvp} and (iii) $W_R \rightarrow \ell \nu_R\rightarrow \ell \ell W_R^* \rightarrow 2\ell 2j$ \cite{Sirunyan:2018pom}, the third one being relevant only in the kinematic regions with $M_{\nu_R} < M_{W_R}$.  The Majorana nature of the right-handed neutrino allows
for probing both the same-sign and opposite-sign dilepton channels~\cite{%
Keung:1983uu}.  Both ATLAS and CMS, in their latest analyses with 37 fb$^{-1}$  and 35.9  fb$^{-1}$ luminosities, respectively, at $\sqrt{s}=13$ TeV, have provided stringent limits on $M_{W_R}$, in the  3.5-4.4  TeV region. CMS \cite{Sirunyan:2017vkm} considered both scenarios of (i) $M_{\nu_R} \ge M_{W_R}$ and (ii) $M_{\nu_R} < M_{W_R}$, where in the second case, in addition to the hadronic decay, $W_R$ can decay to $\nu_R \ell$. The branching fraction of $W_R \rightarrow t\bar b$ is quoted as (0.32 - 0.33) in the first scenario and (0.24 - 0.25) in the second scenario, for a $W_R$ mass range of 500 GeV to 5 TeV, where the 1\% difference is due the phase space effects coming from the top quark mass (as compared to the vanishing light quark mass).  The ATLAS analysis  \cite{Aaboud:2018spl} also considered both scenarios. However, they searched for $W_R$ gauge bosons in final states with two charged leptons and two jets.  Both CMS and ATLAS assumed strict $LR$ symmetry setting $g_R=g_L$ and $V_{\rm CKM}^L = V_{\rm CKM}^R$ in their analyses.
  
%%%%%%%%%%%%%%%%%%%%%%%%%%%%%%%%%%%%%%%%%%%%%%%%%%%%%%%%%
\subsection{Doubly-charged Higgs bosons}
\label{subsec:dch}
%%%%%%%%%%%%%%%%%%%%%%%%%%%%%%%%%%%%%%%%%%%%%%%%%%%%%%%%%
Mass bounds on the doubly-charged Higgs bosons must satisfy consistency with
the various LHC limits. The restrictions on the mass of the left-handed doubly charged Higgs boson $\delta_L^{++}$ are considerably weakened as these can decay into both same sign lepton pairs and same sign $W_L$ pairs. The latter decay mode is not available to the right-handed doubly-charged Higgs boson $\delta_R^{++}$, so the restrictions coming from the doubly-charged Higgs bosons decaying into two same-sign leptons, which are very strong, apply. The exception is the case
in which the doubly-charged Higgs boson decays into a di-tau
final-state~\cite{Chatrchyan:2012ya,CMS:2016cpz}.  Without any significant excess of events, the LHC analyses mentioned presently
provide stringent constraints from direct searches, which require the masses of the doubly
charged scalars to be above   $\sim 600$ GeV ($\sim 500$ GeV for decays into di-taus only).

Not  all constraints coming from (first and second family) dilepton decays can be evaded, as  the right-handed triplet Higgs field are needed to generate masses for the
right-handed neutrinos. In our scenario, $W_R$ production and decays are not affected by the doubly-charged triplet Higgs bosons, and thus the latter can be heavy, ${\cal O}(v_R)$.
%%%%%%%%%%%%%%%%%%%%%%%%%%%%%%%%%%%%%%%
\subsection{Right-handed neutrinos}
\label{subsec:nuR}
%%%%%%%%%%%%%%%%%%%%%%%%%%%%%%%%%%%%%%%
Generic searches for right-handed neutrinos were performed at LEP~\cite{%
Achard:2001qw}, leading to bounds on right-handed Majorana neutrino masses of at
most $90.7$~GeV. In our model the right-handed neutrino mass matrix reads
\begin{equation}
 M_{\nu_{R}}^{ij}=  h_{R}^{ij}v_{R}\ ,
\end{equation}
where $ h^{ij}_{R}$ also dictates the different doubly-charged Higgs branching
ratios. If we enforce that the $SU(2)_{R}$ doubly-charged Higgs boson 
decays mainly into taus, the right-handed tau neutrino turns out to be
significantly heavier than the others. The mass of right-handed neutrinos are determined by  the choices of $v_{R}$ and $h_R$ and can be chosen to be either $M_{\nu_R} \ge M_{W_R}$ or $M_{\nu_R} \le M_{W_R}$, for the two scenarios that we investigate.
We now proceed to analyze the parameter space of the model, and subject it to constraints from phenomenology.

\section{Model Implementation and Constraints}
\label{sec:scan}
In most analyses, $W_R$  bosons are expected to be heavy. 
However all of the analyses assume that the model is manifestly left-right symmetric, that is, the coupling constants are the same for the left and right gauge sectors,  $g_L=g_R$ {\it and} that the quark mass mixing in the right-handed sector $V^R_{CKM}$ is either diagonal, or equal to the one in the left-handed sector (the Cabibbo-Kobayashi-Maskawa matrix, $V^L_{CKM}$). This does not have to be the case, and analyses of a more general model, the so-called asymmetric left-right model exist \cite{Langacker:1989xa,Frank:2010cj}. 

We take this general approach here. We calculate the production cross section and decays of the $W_R$ bosons in the LRSM with $g_L\ne g_R$ and allowing for general entries in the mixing matrix for the right-handed quarks,  $V^R_{CKM}$, parametrized as the left-handed matrix, but allowing the elements to vary independently:
\begin{equation}
V_{\rm CKM}^R =  \begin{bmatrix}
c^R_{12}c^R_{13} & s^R_{12}c^R_{13} & s^R_{13}e^{i\delta_R} \\
-s^R_{12}c^R_{23}-c^R_{12}s^R_{23}s_{13}e^{i\delta_R} & c^R_{12}c^R_{23}-s^R_{12}s^R_{23}s^R_{13}e^{i\delta_R} & s^R_{23}c^R_{13} \\
s^R_{12}s^R_{23}-c^R_{12}c^R_{23}s^R_{13}e^{i\delta_R} & -c^R_{12}c^R_{23}-s^R_{12}c^R_{23}s^R_{13}e^{i\delta_R} & c^R_{23}c^R_{13} 
\end{bmatrix}  
\end{equation}
  We then proceed as follows. 
\begin{itemize}
\item We first choose one value for $M_{W_R}$. Then we vary the parameters $c^R_{12}$, $c^R_{13}$ and $c^R_{23}$ in the range $[-1, 1]$. The phase $\delta_R$ is set to zero (as we are not concerned with CP violation), and we impose matrix  unitary condition. For each set of the randomly chosen $V_{CKM}^R$ parameters as above, we impose the theoretical and experimental constraints including the 
mass bounds and flavor constraints from $K$ and $B$ mesons, as listed in  \autoref{tab:constraints}. This ensures, for instance, that the non-SM neutral bidoublet Higgs boson is very heavy ($ > 10 $ TeV), as required to suppress flavor-violating effects. 
\item  For each value of $M_{W_R}$ we obtain many solutions for $V_{CKM}^R$ consistent with all bounds. Of these solutions, we choose the one yielding the smallest branching ratio for $W_R \to t \bar{b}$ \footnote{We have also tried to minimize $BR(W_R \to jj)$ but found that this choice yields a more restrictive lower mass bound for $W_R$.}.
\item 
%Although solutions are much easier to obtain for higher $M_{W_R}$ masses, the values obtained for the elements of $V^R_{CKM}$ fall into a range for each case. 
Noting that the flavor bounds are very sensitive to these elements,  we fit them carefully for each solution. The $V_{CKM}^R$ elements thus obtained are restricted to a viable ranges of values. We distinguish between two possibilities: one when the $W_R$ mass is lighter than that of the right-handed neutrinos $\nu_R$, and the case where the right-handed neutrino is lighter. In the second case, the right-handed gauge boson can decay also into $\nu_R \ell$, modifying its branching ratio to top and bottom quarks.
For the case of $M_{\nu_R}< M_{W_R} $, these ranges of elements in the right-handed CKM matrix are
\begin{equation}
V_{\rm CKM}^R =  \begin{bmatrix}
[3.63 \times 10^{-3}  -  0.736] & [0.650  -  0.999] & [3.18\times 10^{-2}   -   0.754] \\
[0. 671   -   0.999] &  [1.93\times10^{-3}   -   0.550] & [2.24\times 10^{-2}   -   0. 501] \\
[2.05\times 10^{-4}   -   0.439] &  [3.01\times10^{-2}   -   0. 619] &  [0.781   -   0.996]  
\end{bmatrix},  
\end{equation}

while in case of $M_{\nu_R}> M_{W_R} $ the ranges of elements in the right-handed CKM matrix are:
\begin{equation}
V_{\rm CKM}^R =  \begin{bmatrix}
[1.28 \times 10^{-3}  -  9.91\times10^{-2}] & [0.858  -  0.996] & [5.25\times 10^{-2}   -   0.504] \\
[0. 805   -   0.997] &  [8.68\times10^{-5}   -   5.22\times 10^{-2}] & [4.16\times 10^{-4}   -   0. 585] \\
[9.30\times 10^{-3}   -   0.589] &  [2.62\times10^{-2}   -   0. 511] &  [0.807   -   0.998]  
\end{bmatrix}.  
\end{equation}

\item When $M_{\nu_R} > M_{W_R}$, solutions emerge allowing for low values of $BR(W_R \to t \bar{b})$, which vary from about 23.3\% for high $M_{W_R}$, to about 29\% for low $M_{W_R}$, while when $M_{\nu_R} < M_{W_R}$, this ratio changes from 15.7\% for high $M_{W_R}$, to about 24.7\% for low $M_{W_R}$ .
\end{itemize}

\begin{table}{
		\setlength\tabcolsep{6pt}
		\renewcommand{\arraystretch}{2.0}
		\begin{tabular}{|l|c|c||l|c|c|}
			\hline
			Observable & Constraints & Ref. & Observable & Constraints & Ref.\\
			\hline
			$ \Delta{B_s} $              & [10.2-26.4] & \cite{Jubb:2016mvq} &
						$ \Delta{B_d} $                   & [0.294-0.762] & \cite{Jubb:2016mvq} \\
			$ \Delta{M_K} $          & $<$ 5.00 $\times 10^{-15}$  & \cite{PhysRevLett.102.211802} & 
						$ \frac{\Delta{M_K}}{\Delta{M_K^{SM}}} $    &  [0.7-1.3] & \cite{PhysRevLett.102.211802}   \\
			$ \epsilon_K $    & $<$ 3.00 $\times 10^{-3}$   & \cite{PhysRevLett.102.211802}   & 
						$\frac{\epsilon_K}{\epsilon^{SM}_K} $   & [0.7-1.3] & \cite{PhysRevLett.102.211802} \\						
			BR$(B^0 \to X_s \gamma) $  & $  [2.99,3.87]\times10^{-4} $ & \cite{Amhis:2012bh} &
						$\frac{BR(B^0 \to X_s \gamma)}{BR(B^0 \to X_s \gamma)_{SM}} $    &  [0.7-1.3] & \cite{Amhis:2012bh} \\
			$M_{h} $ & $ [124,126] $ GeV                                & \cite{Chatrchyan:2012xdj} &
						$M_{H_{1,2}^{\pm\pm}} $            & $>$ 535 GeV & \cite{Aaboud:2018qcu, CMS:2017pet} \\
			$M_{H_4,A_2,H_2^{\pm}} $ &  $> 4.75 \times M_{W_R}$                                & \cite{Bonilla:2016fqd} &
						            &      & -\\								
			\hline 
		\end{tabular}
		\caption{\label{tab:constraints} Current experimental bounds imposed for consistent solutions.}}
	\end{table}

For this analysis, a version of an LRSM model produced with {\sc SARAH 4.13.0} \cite{Staub:2008uz,Staub:2010jh,Staub:2015kfa} was implemented in {\sc SPHENO} 4.0.3 package \cite{Porod:2003um,Porod:2011nf}. We use {\sc HiggsBounds} \cite{Bechtle:2013wla} and {\sc HiggsSignals} \cite{Bechtle:2013xfa} to test out the signal strengths of the SM-like Higgs state and check the consistency of the Higgs sector for each solution. The relevant cross sections were calculated by {\sc MG5\_aMC@NLO v2.6.3.2} \cite{Alwall:2011uj,Alwall:2014hca} using the NNPDF2.3 \cite{Ball:2012cx} parton distribution function (PDF) set. In order to get cross sections at the next-to-leading order (NLO) accuracy, a version of an LRSM model was produced with {\sc FeynRules} \cite{Alloul:2013bka}. An UFO \cite{Degrande:2011ua} file obtained from {\sc FeynRules} is implemented in {\sc MG5\_aMC@NLO} \cite{Roitgrund:2014zka} and then used for the numerical evaluation of the hard-scattering matrix elements. We use {\sc pySLHA 3.2.2} package \cite{Buckley:2013jua} for manipulating SUSY Les Houches Accord (SLHA) files during the numerical analysis performed in this work.

%An UFO \cite{Degrande:2011ua} file containing tree-level vertices was obtained from SARAH and then used by MG5\_aMC@NLO for the numerical evaluation of the hard-scattering matrix elements. In order to be more accurate in calculating the $W_R$ production cross section, we multiplied the production cross section with a $K$-factor in the W$_R \to tb$ and W$_R \to jj$ channels, which accounts for the next-to-leading order (NLO) corrections in QCD. The relevant $K$-factor was calculated for some benchmarks dividing the NLO cross section by the cross section value at the LO level. To obtain NLO cross sections, a version of an LRSM model produced with {\sc FeynRules} \cite{Alloul:2013bka} was also implemented in {\sc MG5\_aMC@NLO} \cite{Roitgrund:2014zka}. Finally, the cross section obtained by {\sc FeynRules} implementation at the NLO was divided by the cross section value obtained by {\sc SARAH} implementation at the LO level, and the corresponding $K$-factor was obtained. Since the $K$-factors obtained from some benchmarks are close but not exactly the same, we created a $K$-factor function for each exclusion channel, which  varies along $W_R$ mass range. $W_R$ production cross sections  were then calculated multiplying the cross section at the LO with these $K$-factor functions. After we obtained the solutions, we also use {\sc HiggsBounds} \cite{Bechtle:2013wla} and {\sc HiggsSignals} \cite{Bechtle:2013xfa} to test out and compare the consistency of the Higgs sector of the solutions.

We performed random scans over the parameter space, as illustrated in Table \ref{tab:scan_lim}, and imposed the mass bounds on all the particles, as well as other constraints as given in Table \ref{tab:constraints}. In scanning the parameter space, we used the interface which employs the Metropolis-Hasting algorithm. In what follows,  we shall distinguish between two cases, Scenario I, $ M_{\nu_R} >M_{W_R} $, and Scenario II, $ M_{\nu_R} < M_{W_R} $.

\begin{table}
	\setlength\tabcolsep{20pt}
	\renewcommand{\arraystretch}{2.0}
	\begin{tabular}{|c|c||c|c|}
		\hline
		Parameter      & Scanned range& Parameter      & Scanned range\\
		\hline
		$v_{R}$          & $[2.2, ~20]$~TeV   &  diag$(h_R^{ij})$   & $[0.001,~ 1]$ \\
		$V_{\rm CKM}^{R}$:~~$c^R_{12},~c^R_{13},~c^R_{23}$          & $[-1,~ 1]$   &    & \\
		\hline
	\end{tabular}
	$\vspace{0.5cm}$
	\caption{\label{tab:scan_lim} Scanned parameter space.}
\end{table}

%%%%%%%%%%%%%%%%%%%%%%%%%%%%%%%%%%%%%%%%%%%%%%%%%%%%%
\section{$M_{\rm W_R} $ Mass Bounds: the analysis}
\label{sec:analysis}
%%%%%%%%%%%%%%%%%%%%%%%%%%%%%%%%%%%%%%%%%%%%%%%%%%%%%
In a general left-right model, the masses of the $Z_R$ and $W_R$ gauge bosons are related, but the mass ratios  depend sensitively on the values of the coupling constants $g_R$ and $g_{B-L}$. While the mass of $W_R$ is proportional to $g_R$, the mass of the $Z_R$ boson is proportional to $\sqrt{g_R^2+g_{B-L}^2}$. Breaking the symmetry to $U(1)_{\rm em}$ imposes  that couplings constants  are related through
\begin{equation}
\frac{1}{e^2}=\frac{1}{g_L^2}+ \frac{1}{g_R^2}+\frac{1}{g_{B-L}^2}\, ,
\end{equation}
and therefore decreasing $g_R$ results in increasing $g_{B-L}$. Note also that these couplings are related, through $SU(2)_R \otimes U(1)_{B-L} \to U(1)_Y$ symmetry breaking to the coupling of the hypercharge group, $g_Y$, through
\begin{equation}
\frac{1}{g_Y^2}= \frac{1}{g_R^2}+\frac{1}{g_{B-L}^2}\, .
\end{equation}
This means that, setting $\displaystyle \sin \phi=\frac{g_{B-L}}{\sqrt{g_R^2+g_{B-L}^2}}$, and with the usual definition of $\displaystyle \sin \theta_W=\frac{g_{Y}}{\sqrt{g_L^2+g_{Y}^2}}$, we obtain
\begin{equation}
\tan \theta_W=\frac{g_R \sin \phi}{g_L} \le\frac{g_R}{g_L} \, ,
\end{equation}
showing clearly that we cannot lower $g_R$ below its minimum value, $g_L \tan \theta_W$. Lowering $g_R$ will decrease the production cross section for $W_R$, while having no effect on its branching ratios. We analyze the case where $g_R=g_L$, as well as when $g_L \neq g_R=0.37$, its allowed minimal value. 

In the top left panel of \autoref{fig:LRSMheavyneutrino} we plot the relationship between the two gauge boson masses, $W_R$ and $Z_R$, for the two cases ($g_R=g_L$ and $g_L \neq g_R=0.37$). As seen from the figure, the $Z_R$ mass is much closer to $W_R$ mass when $g_R=g_L$, and in that case the $Z_R$ mass bounds (especially coming from  production, followed by decays into dileptons) may have an effect on the $W_R$ mass bounds. While for $g_L \neq g_R=0.37$, $Z_R$ is expected to be much heavier than $W_R$, making the latter more likely to be much lighter and more likely to be the first observed.

%%%%%%%%%%%%%%%%%%%%%%%%%%%%%%%%%%%%%%%%%%%%%%%%%%%%%
\subsection{Scenario I: $M_{\nu_R} > M_{W_R} $}
\label{subsec:LRSMheavyneutrino}
%%%%%%%%%%%%%%%%%%%%%%%%%%%%%%%%%%%%%%%%%%%%%%%%%%%%%
We now proceed to investigate the case where the right handed neutrino is heavier than $W_R$, (so the on-shell decay $W_R \to \nu_R \ell$ is disallowed) and where the possible decay channels for $W_R$ are:
$$W_R \to jj (q \bar{q}^\prime),  \,\, W_R \to W_L h, \, \, W_R \to W_L Z \, \, {\rm and} \, \, W_R \to W_L hh\, ,$$
all the other Higgs states being very heavy. 
Of these, the three-body decay $W_L hh$ is very weak with $\frac{\Gamma(W_R\to W_L h)}{\Gamma(W_R\to W_L hh)} \sim 1/v$, while the decay $W_R\to W_L h$ depends on $\tan \beta$.  Most analyses assume $\tan \beta$ to be very small ($\sim 0.01$), yielding  $BR(W_R\to W_L h)$ to be negligible. Perturbativity bounds alone require $\tan \beta < 0.8$, however the mass of the SM-like Higgs boson $h$ also depends on $\tan \beta$, and values of $\tan\beta >0.6$ result in the instability of the $h$ mass. To keep our analysis general, we investigated the production and decays of the $W_R$ mass for two cases: small $\tan \beta=0.01$, and large $\tan\beta=0.5$. In addition, we allow for two values of $g_R$, viz. $g_R=g_L$, and $g_L \neq g_R=0.37$, as well as vary matrix elements of $V^R_{\rm CKM}$ to show how the results are affected. The resulting plots for $\sigma (pp \to W_R) \times BR(W_R \to t \bar{b})$ are given in the top right plane of  \autoref{fig:LRSMheavyneutrino}, where we compare our four different cases (different colour-coded, as indicated in the panel insertion) with the CMS result \cite{Sirunyan:2017vkm} using collision data collected at $\sqrt{s}=$13 TeV with ${\cal L}$=35.9 fb$^{-1}$.  We show the observed and expected limit curves for the combined electron and muon final states. For $g_L=g_R$,  $V_{\rm CKM}^L= V_{\rm CKM}^R$ and $\tan \beta=0.01$, we note that the branching ratio $W_R$ into $t{\bar b}$ ranges from 32\% to 33\%, as $W_R$  decays into $q{\bar q}^\prime$ pairs democratically. We confirm the CMS result \cite{Sirunyan:2017vkm} and exclude $W_R$ boson mass up to 3.6 TeV. For the  case where $g_R \neq g_L$ and $V_{\rm CKM}^L=V_{\rm CKM}^R$, represented by the blue line,  we set $g_R \simeq 0.37$ and $\tan \beta$ = 0.01. The $W_R$ production cross section decreases due to relatively small $g_R$, and the exclusion limit for $W_R$ masses can be reduced to 2.7 TeV in that scenario. Increasing  $\tan\beta$ to 0.5 and $g_R \simeq 0.37$, enhances the branching ratios of $W_R \to W_L h$  to about $\sim$ 1.95\% and the branching ratio of $W_R \to W_L Z_L$ to about $\sim$ 2.0\%. In this case, the branching ratio of $W_R \to t {\bar b}$ is reduced slightly, to  31.0\% - 31.8\%, as shown by the pink line.   As can be read from the plot, this reduces the $W_R$ mass limits only slightly,  to 2675 GeV. However, when we allow $V_{\rm CKM}^L \ne V_{\rm CKM}^R$, $\tan\beta$= 0.5 and $g_R \simeq 0.37$, this maximizes decays of $W_R$ into other final states, and the branching ratio of $W_R \to t{\bar b}$ is reduced substantially: from about 20\% for high $M_{W_R}$ (4 TeV) to about 29\% for low $M_{W_R}$ (1.5 TeV). The orange line in the top right plane of  \autoref{fig:LRSMheavyneutrino} represents our result for this scenario, and the exclusion limit is reduced to 2360 GeV with respect to observed limit, whereas it can be estimated at 1940 GeV based on the expected limit.

In the left bottom panel of \autoref{fig:LRSMheavyneutrino} we plot the cross section of $pp\rightarrow W_R \to jj$ vs $W_R$ mass, and compare it to the ATLAS result \cite{Aaboud:2017yvp}  at $\sqrt{s}=$13 TeV for ${\cal L}$=37 fb$^{-1}$. (For comparison, we included their acceptance factor $A$ ). The red curve represents the exclusion limit for $W_R$ mass when the gauge couplings $g_L=g_R$,  and $\tan\beta$ = 0.01. The branching fraction of $W_R \to jj$ varies slightly with mass. We keep the same color coding for curves as in the previous panel. The mass restrictions are comparable, but slightly weaker than those for the $W_R \to t \bar{b}$ decay, ranging from $M_{W_R} \ge$ 3625 GeV when $g_L=g_R$, $\tan \beta=0.01$  and $V^R_{\rm CKM}=V^L_{\rm CKM}$, to  $M_{W_R} \ge 2.0$ TeV when $g_R=0.37$, $\tan \beta=0.5$ and $V^R_{\rm CKM}\ne V^L_{\rm CKM}$. Neither  results are particularly sensitive to values of $\tan \beta$, but depend on choices for $g_R$ and $V^R_{\rm CKM}$.

\begin{figure}	
	$\begin{array}{cc}
	\includegraphics[width=.48\columnwidth]{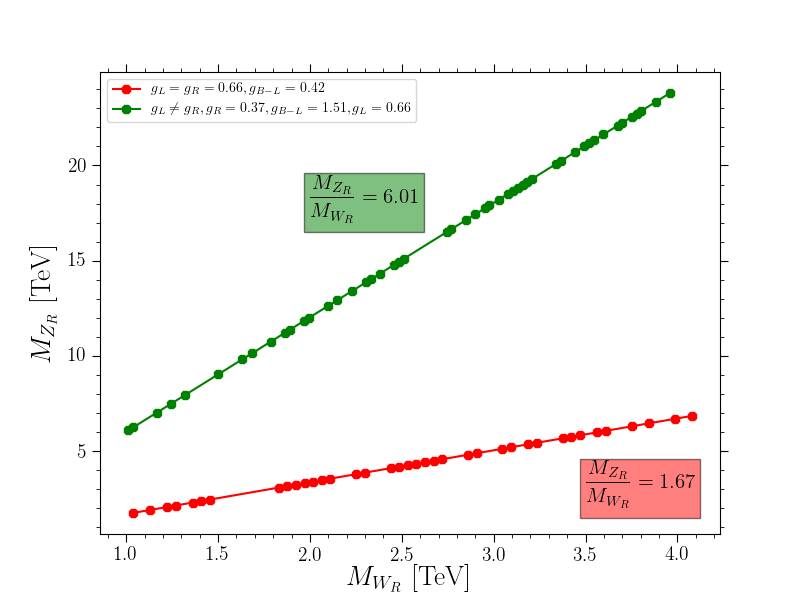}
	\includegraphics[width=.48\columnwidth]{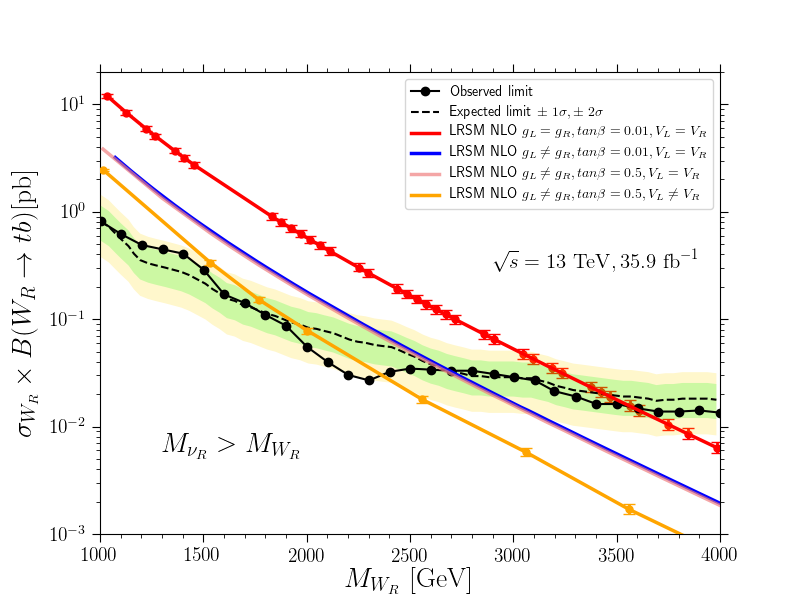}\\
	\includegraphics[width=.48\columnwidth]{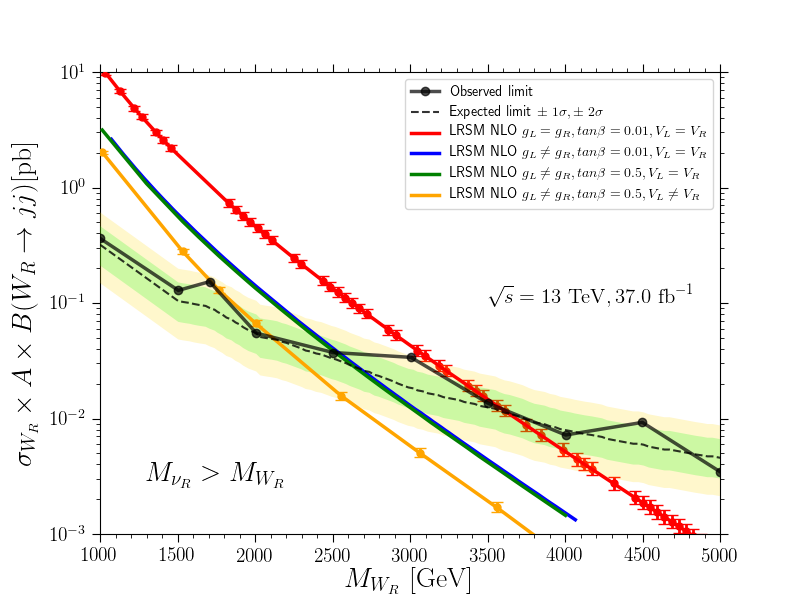}
	\includegraphics[width=.48\columnwidth]{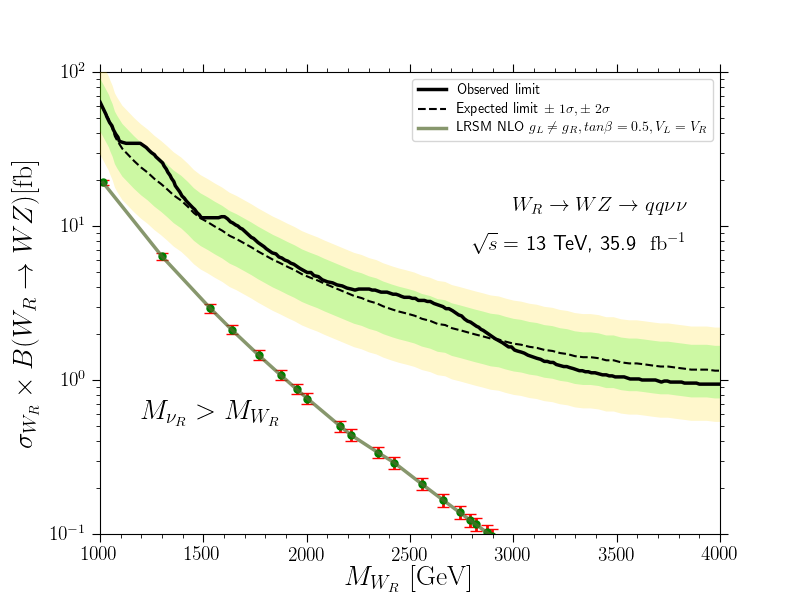}
	\end{array}$
	\caption{ (Top left): The comparison of $W_R $  and $Z_R$ masses for $g_L=g_R$ and $g_L \ne g_R=0.36$; (top right) The cross section of W$_R \to tb$ vs $W_R$ mass for different values of $\tan \beta$ and $g_R$; (bottom left)  The cross section of $W_R \to jj$ vs $W_R$ mass, for different values of $\tan \beta$ and $g_R$, compared to ATLAS data \cite{Aaboud:2017yvp}; (bottom right) The cross section of $W_R \to W_L Z$ vs $W_R$ mass, where the experimental data shown for comparison is from CMS \cite{Sirunyan:2018ivv}.  The last three plots are for the case when $M_{\nu_R}>M_{W_R}$.}
\label{fig:LRSMheavyneutrino}
\end{figure}

The right bottom plane of \autoref{fig:LRSMheavyneutrino} represents comparison with the CMS result \cite{Sirunyan:2018ivv} which searches for decay $W_R \to  W_L Z$  in the $\nu \nu q {\bar q}^\prime$ final state using the $pp$ collision data collected at $\sqrt{s}=$13 TeV at ${\cal L}$=35.9 fb$^{-1}$. Since the branching ratio of $W_R \to W_L Z$ is negligible ($< 10^{-4}$) for small $\tan \beta$ values, we show our result in this channel only for  large $\tan\beta$ values (i.e. $\tan\beta$ = 0.5), when the branching ratio of $W_R \to W_L Z$ is at about 2 $\%$. As seen from the graph, our  curves are always below the experimental curves and thus  $W_R$ masses cannot be excluded using the $W_R \to W_L Z \to q {\bar q}^\prime \nu \nu$ channel. 

%\newpage

%%%%%%%%%%%%%%%%%%%%%%%%%%%%%%%%%%%%%%%%%%%%%%%%%%%%%
\subsection{Scenario II: $ M_{\nu_R} < M_{W_R} $}
\label{subsec:LRSMlightneutrino}
%%%%%%%%%%%%%%%%%%%%%%%%%%%%%%%%%%%%%%%%%%%%%%%%%%%%%
The case where the $W_R$ is heavier than the right-handed neutrino opens the possibility of $W_R \to \nu_R \ell$, followed by decay of  $\nu_R\rightarrow \ell W_R^*\rightarrow \ell jj$ giving rise to an $\ell \ell jj$ signature. The Majorana nature of the right-handed neutrino allows for probing both the same-sign and opposite-sign dilepton channels \cite{Keung:1983uu}. Both the ATLAS \cite{Aaboud:2018spl} and CMS \cite{Sirunyan:2018pom} collaborations have looked for such a $W_R$ signal, excluding $W_R$ masses up to about 4.7 TeV for right-handed (muon or electron) neutrino masses up to  $3.1$ TeV~\cite{Sirunyan:2018pom}. For lower right-handed neutrino masses (below $200$~GeV), the bound is less restrictive than the one originating from dijet searches. For the tau channel, it is even much weaker, with $M_{W_R}$ constrained to be only smaller than $2.9$~TeV~\cite{Sirunyan:2017yrk}.

In  \autoref{fig:LRSMneuthalf}, we plot our results and compare it with the current experimental results. Throughout our analysis we  assign one right-handed neutrino mass to be half of the mass of $W_R$, for comparison with the experimental analysis. (The results obtained though are typical for large gaps between $\nu_R$, $W_R$ masses.) The top left plane of \autoref{fig:LRSMneuthalf} shows the exclusion based on $W_R \to t{\bar b}$ channel. The red line shows the result for $g_L = g_R$, $\tan \beta = 0.01$ and $V_{\rm CKM}^L = V_{\rm CKM}^R$. In addition to the decays into hadrons, $W_R$ also decays into $\nu _R\ell$,  about 5.8\% for each lepton flavor. We note that the branching ratio of $W_R$ into $t {\bar b}$ ranges from 26.5\% to 27.3\%. We confirmed the CMS result \cite{Sirunyan:2017vkm} and find out that the observed (expected) 95\% confidence level (CL) lower limit for $W_R$ mass bounds is 3450 (3320) GeV. We also study the effect of $g_R\neq g_L$ by setting $g_R \simeq 0.37$ and $\tan \beta = 0.01$. In this scenario the leptonic  decay rates ($W_R \to \nu_R\ell$) increase 
 to about 6.7\% for each lepton flavor, and the branching ratio of $W_R \to t{\bar b}$ ranges from 25.7\% to 26.5\%. The estimated observed (expected) bound on $W_R$ masses is 2575 (2375) GeV at 95\% CL, as seen from the blue line. In order to see the effect of large $\tan \beta$ and $g_R \neq g_L$, we set $g_R \simeq 0.37$ and $\tan\beta = 0.5$. In addition to $W_R$ decays into $t{\bar b}$, $q{\bar q}^\prime$ and $\nu_R\ell$, the decays of $W_R \to   h W_L$ and $W_R \to W_L Z$ are non-negligible, with each one at about 2\% BR. Therefore, the branching ratio of $W_R$ into $t{\bar b}$ is reduced slightly further,  and now ranges between 24.8\% and 25.6\%. The green line shows our result, and the estimated observed (expected) limit at 95\% CL is 2565 (2350) GeV. Finally, setting $g_R \simeq 0.37$, $\tan\beta$ = 0.5, and varying the matrix elements of $V_{\rm CKM}^R$, the branching ratio of $W_R \to t{\bar b}$ varies from about 15.7\% for high $M_{W_R}$ (4 TeV), to about 24.7\% for low $M_{W_R}$ (1.5 TeV). The pink line shows our result, and the estimated observed (expected) limit at 95\% CL is 2320 (1850) GeV.

The top right plane of \autoref{fig:LRSMneuthalf} shows the $W_R$ mass exclusion based on $W_R \to jj$ channel. The estimated exclusion limits vary from 3.5 TeV for the scenario where $g_L = g_R$,  $\tan\beta$ = 0.01 and $V_{\rm CKM}^L = V_{\rm CKM}^R$, to 2.0 TeV for the scenario which we set $g_L \neq g_R=0.37$,  $\tan \beta$ = 0.5 and $V_{\rm CKM}^L \ne V_{\rm CKM}^R$ as can be read in detail from  \autoref{tab:ExclusionBoundsII}. 

 In addition to $W_R \to t {\bar b}$ and $W_R \to jj$ channels, the most stringent bounds come from the channel in which the $W_R$ boson decays to a first or second generation charged lepton and a heavy neutrino of the same lepton flavor. Both ATLAS \cite{Aaboud:2018spl} and CMS \cite{Sirunyan:2018pom} assume that the heavy neutrino further decays to another charged lepton of the same flavor and a virtual $W_R^\star$ with a 100\% rate. The virtual $W_R^\star$ then decays into two light quarks, producing the decay chain

\begin{equation}
W_R \to \ell \nu_R \to \ell \ell W_R^\star \to \ell \ell q q^\prime, ~~\ell = e \hspace{0.15cm} {\rm or} \hspace{0.15cm}  \mu \, .
\end{equation}

However,  this is true only for small $\tan\beta$ values, where the corresponding mixing angle $\xi$ between the two charged gauge bosons  is extremely small. In the large  $\tan \beta$ case, the mixing between  $W_L$ and $W_R$, although small enough to satisfy flavor and CP bounds, becomes important, inducing a $W_L$  contribution to the above channel, producing the decay chain 
 
\begin{equation}
W_R \to \ell \nu_R \to \ell \ell W_L \to \ell \ell q q^\prime,~~ \ell = e \hspace{0.15cm} {\rm or} \hspace{0.15cm}  \mu \,.
\end{equation} 

This can be traced to  the interaction of right-handed neutrino field with $W_L^{+\mu}\ell$ \cite{Roitgrund:2014zka}:
%
%\begin{equation}
%\overline{\nu}W_L^{+\mu}\ell \longrightarrow \frac{i}{\sqrt{2}} \gamma^\mu \left(g_L P_L K_L\cos\xi -g_R P_R K_R\sin \xi \right) \,  
%\label{eq:interactionNRWLel}
%\end{equation}
%\begin{equation}
%\overline{\nu}W_R^{+\mu}\ell \longrightarrow \frac{i}{\sqrt{2}} \gamma^\mu \left(g_R P_R K_R\cos\xi -g_L P_L K_L\sin \xi \right) \,
%\label{eq:interactionNRWRel}
%\end{equation}
%
%
\begin{eqnarray}
\label{eq:interactionNRWLel}
\overline{\nu}W_L^{+\mu}\ell &\longrightarrow& \frac{i}{\sqrt{2}} \gamma^\mu \left(g_L P_L K_L\cos\xi -g_R P_R K_R\sin \xi \right) \, , {\rm in~addition~ to~ the~ usual~ interaction~ with~ }W_R \\
\label{eq:interactionNRWRel}
\overline{\nu}W_R^{+\mu}\ell &\longrightarrow& \frac{i}{\sqrt{2}} \gamma^\mu \left(g_R P_R K_R\cos\xi -g_L P_L K_L\sin \xi \right) \,
\end{eqnarray}

where $K_L$ and $K_R$ are PMNS mixing matrices in the left and right leptonic sectors, defined as
\begin{equation}
K_L=V_L^{\nu\dagger}V_L^\ell, \hspace{4.0cm} K_R=V_R^{\nu\dagger}V_R^\ell.
\end{equation} 
%%%%%%%%%%%%%%%%%Nov 22 up to here%%%%%%%%%%%%%%%%%%%%%%%%%%%%%%%%%%%%%%%%%%%%%%
%
Remember that, while we denote the gauge bosons by $W_{R,L}$, in the presence of mixing, each physical eigenstate is a mixture of the the left- and right-handed gauge bosons. 
For the case of no mixing between charged gauge bosons, $W_L$ is purely left-handed, and consequently, $\nu_R$ will not decay to this state. 
%the whole contribution to this decay is obtained through the first term in \autoref{eq:interactionNRWLel}, and it is negligible. 
However, the second term in \autoref{eq:interactionNRWLel}  is important  as the mixing $\xi$ increases, and $M_{\nu_R}$ decreases. Contributions to $\ell \ell jj$ final states through $W_R^\star$ and $W_L$ are illustrated in the \autoref{fig:WLWRmixing}, where we set $M_{W_R} =$ 3.5 TeV both in the left and right plane. The green line shows the branching ratio of $\nu_R \to \ell qq^\prime$, namely, the contribution to $\ell \ell jj$ final states through virtual $W_R^\star$ boson, \autoref{eq:interactionNRWRel},  while the red curve represents the branching ratio of $\nu_R \to W_L \ell$, namely, the contribution to $\ell \ell jj$ final state through an $W_L$ boson \autoref{eq:interactionNRWLel}. The lower $x$-axis and the top $x$-axis represent the mixing angle $\xi$ between the charged gauge bosons and  $\tan \beta$, respectively,  thus showing the correlation between them. In the left plot, we show this correlation for the case of $g_L =  g_R$, while the right plot depicts the same correlation, but for the case  $g_L \neq g_R=0.37$. The plots indicate that,  when both $\tan \beta$ and $\xi$ are small, all contributions proceed through a virtual $W_R^\star$ boson.   As an increase in  $\tan \beta$ induces an increase in $\xi$, the contributions to $\ell \ell jj$ final states receive contributions through both virtual $W_R^\star$ and real $W_L$ bosons. When $g_L = g_R$, $\tan \beta = 0.01$ and $M_{W_R} =3.5$ TeV, approximately 98\% of the total contribution to $\ell \ell jj$ final state comes through the virtual $W_R^\star$ boson, with  $W_L$ boson contribution to about 2\%. The mixing between two charged gauge bosons become more important in case of $g_L \neq g_R$. When $g_L \neq  g_R=0.37$,  $\tan \beta = 0.5$ and $M_{W_R} =3.5$  TeV, the contribution to $\ell \ell jj $ final state through $W_L$ boson increases to 30\%, meaning that the contribution through a virtual $W_R^\star$ boson has been reduced to 70\%\footnote{The values chosen for $\tan \beta$ and mixing angle $\xi$ satisfy all flavor constraints in \autoref{tab:constraints}. }.  Since we increased the ratio of $g_L/g_R$ from 1.0 to 1.79 by setting $g_R \simeq 0.37$,  the vacuum expectation value $v_R$ needs to be increased approximately by 1.8 times  to obtain the the same $W_R$ boson mass as in the $g_L=g_R$ case, the corresponding $\xi$ value decreasing slightly compared to the case where $g_L \neq g_R$. The decrease in $g_R$ and increase in $\tan\beta$ increases the contribution through $W_L$ into $\ell \ell  jj$ final state. 
%This effect can be read from \autoref{eq:interactionNRWLel}. 
%For the case of small mixing between charged gauge bosons, the second term in  \autoref{eq:interactionNRWLel} is negligible, and the contribution through the first term is not sufficient. Therefore, 
In case of small $\xi$, which also corresponds to small $\tan \beta$, the branching ratio of $\nu_R \to W_L \ell$ is negligible ($< 10^{-4}$). As the mixing $\xi$ between charged gauge bosons increases (concurrently with $\tan \beta$), the second term in \autoref{eq:interactionNRWLel} starts to give non-negligible contributions, and 
%, as seen in \autoref{eq:interactionNRWLel}. 
as a result, the branching ratio of $\nu_R \to  W _L \ell$ becomes larger.
%, and the contribution to $\ell \ell$jj final state is also obtained through W$_L$ bosons in the scenarios where we use large tan$\beta$. 
In case of $g_R = g_L$ and small $\tan \beta$, the main contribution to $W_R \to \nu_R \ell$ comes from the first term  in \autoref{eq:interactionNRWRel}.
%, as seen in  \autoref{eq:interactionNRWLel}. 
Increasing  $\tan\beta$ and decreasing  $g_R$ slightly reduces the contribution of this term since it is proportional to $g_R \cos\xi$. Therefore, the branching ratio of $W_R \to \nu_R \ell$ slightly decreases in the case of $g_L \neq g_R$, and the contribution to $\ell \ell jj$ final state is compensated  through $W_L$ contributions, slightly more so compared to the $g_L =g_R$ case.

The bottom planes of \autoref{fig:LRSMneuthalf} show the result of the analysis of the $\ell \ell jj$ final states. The bottom left panel plots the result for decays into $\ell \ell=ee$ channel whereas the bottom right one is  for the $\ell \ell= \mu \mu$ channel. For the scenarios where $\tan \beta =0.01$, the contribution through $W_L$ bosons is suppressed. Therefore, the main contribution to $\ell \ell jj$ final states comes via the virtual $W_R^\star$ boson. However, for consistency we sum up the contributions through the virtual $W_R^\star$ and real $W_L$ boson, and our graphs for large $\tan \beta$ values in the bottom planes represent the combined contribution. The most stringent bounds occur for $g_L= g_R$, $\tan \beta  = 0.01$ and $V_{\rm CKM}^L = V_{\rm CKM}^R$. We confirmed in that case the CMS result \cite{Sirunyan:2018pom} and find out that the observed (expected) 95 \% CL lower limit on $W_R$ masses is 4420 (4420) GeV in the $ee$ channel and 4420 (4500) GeV in the $\mu \mu$ channel. 

The observed (expected) limit is reduced to 3800 (3800) GeV in the $ee$ channel and 3800 (3950) GeV in the $\mu \mu$ channel when $g_L \neq g_R$,  $\tan \beta$ = 0.01 and $V_{\rm CKM}^L = V_{\rm CKM}^R$. In the scenarios where $\tan\beta = 0.5$, contributions to $\ell \ell jj $ final states proceed through both virtual $W_R^\star$ and $W_L$ bosons. When the $W_R$ mass is about 1 TeV, approximately 90.5\% of the combined contribution is obtained through $W_L$ bosons, limiting the virtual $W_R^\star$ boson contribution to  9.5\%. However, this relation is flipped when $W_R$ mass is about 4 TeV, where 77.5\% of the combined contribution to $\ell \ell jj $ final states is obtained through the virtual $W_R^\star$ boson, leaving the $W_L$ boson to contribute at  20.6\%. In the scenario where we  $g_L \neq  g _R=0.37$, $\tan \beta= 0.5$,  and $V_{\rm CKM}^L = V_{\rm CKM}^R$, we improve the bounds to where the observed (expected) 95\% CL lower limit is 3725 (3720) GeV in the $ee$ channel, and 3750 (3900) GeV in the $\mu \mu$ channel. In addition to lowering $g_R$ and increasing $\tan \beta$, we verified the effect of different CKM matrices, allowing $V_{\rm CKM}^L \neq V_{\rm CKM}^R$ in our final scenario. The partial contributions through virtual $W_R^\star$ and $W_L$ in this scenario are very close to the one where $V_{\rm CKM}^L = V_{\rm CKM}^R$ and $g_R \neq  g _L$, and $\tan \beta = 0.5$. In this case the results are least constraining and we find  that the observed (expected) 95\% CL lower limit is 3100 (3300) GeV in the $ee$ channel and 3350 (3400) GeV in the $\mu \mu$ channel.
%%%%%%%%%%%%%%%%%%%%%%%%%%%%%%%%%%%%%%%%%%%%%%%%%
\begin{figure}
	$\begin{array}{cc}
	\includegraphics[scale=0.43]{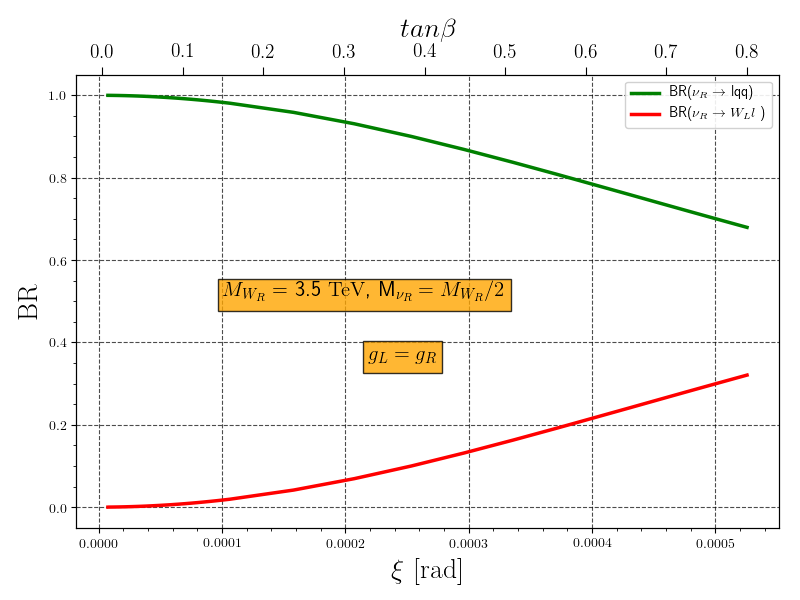}&
	\includegraphics[scale=0.43]{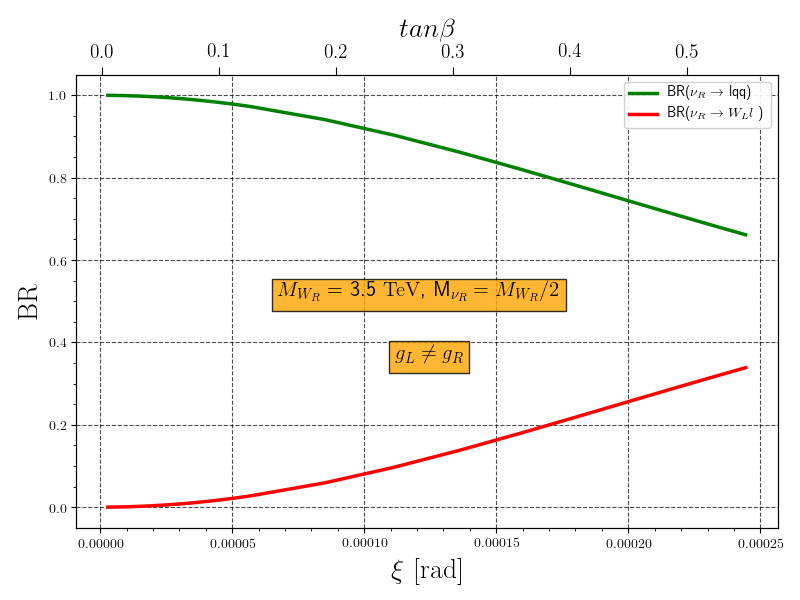}
	\end{array}$
	\caption{ Branching ratios of $\nu_R \to lqq^\prime$ and $\nu_R \to W_L \ell$ versus mixing angle $\xi$ between $W_L$ and $W_R$ (bottom x-axis), changing with tan $\beta$ (upper x-axis) when $g_L = g_R$ (left) and when $g_L \neq g_R=0.37$ (right). We set $V_{\rm CKM}^L =  V_{\rm CKM}^R$ for this analysis.}
	\label{fig:WLWRmixing}
\end{figure}

We now proceed to explore constraints on the $W_R$ and related $\nu_R$ masses and compare these to the CMS and ATLAS analyses. 
%%%%%%%%%%%%%%%%%%%%%%%%%%%%%%%%%%
\subsection{Correlating $W_R$ and $\nu_R$ mass bounds}
\label{subsec:MRnuR}
%%%%%%%%%%%%%%%%%%%%%%%%%%%%%%%%%%%%%%%%%%%%
In  \autoref{fig:LRSMneuthalf_2}, we analyze the correlations  in the two dimensional $M_{W_R} - M_{\nu_R}$ mass plane, covering a  range of neutrino masses both below and above the $W_R$ boson mass. Contrary to the CMS analysis \cite{Sirunyan:2018pom}, which assumes that only one heavy neutrino flavor $\nu_R$ contributes significantly to the $W_R$ decay width, in our analysis all three heavy right-handed neutrino flavors contribute democratically. The $W_R$ production cross section is calculated for each solution in this 2D plane using {\sc MG5\_aMC@NLO},  and the observed (expected) 95\% CL limits obtained from our analysis are applied to explore  excluded regions. The expected and observed upper limits on the cross section for different $W_R$ and $\nu_R$ mass hypotheses are  compared with the latest CMS results \cite{Sirunyan:2018pom} @ ${\cal L}=35.9$ fb$^{-1}$ and ATLAS results \cite{Aaboud:2018spl} @ ${\cal L}=36.1$ fb$^{-1}$, as seen in  \autoref{fig:LRSMneuthalf_2}. Note that we generate our results using the CMS \cite{Sirunyan:2018pom} data, as this is available. The ATLAS analysis, although more recent and at a slightly higher luminosity, is able to rule out a small subset of parameter points in the $M_{W_R} < M_{\nu_R}$ region. However they do not share their observed (expected) cross section plots publicly. Because of that, when we extrapolate our results for slightly higher luminosity in that region, our cross sections are very small, and we do not obtain any restrictions. Thus, we decided it safer to compare our analysis with the existing data points provided by CMS, while indicating restrictions from both experimental analyses on the plots.

In the $M_{\nu_R} < M_{W_R}$ case, we assume that the contribution comes through the following decay chain:
\begin{equation}
W_R \to \ell \nu_R \to \ell \ell W_R^\star \to \ell \ell q q^\prime,~~ \ell = e \hspace{0.15cm} {\rm or} \hspace{0.15cm}  \mu \, , 
\end{equation}
while in the $M_{\nu_R} > M_{W_R}$ case, we assume that the contribution comes through the following decay chain:
\begin{equation}
W_R^\star \to \ell \nu_R \to \ell \ell W_R \to \ell \ell q q^\prime, ~~ \ell = e \hspace{0.15cm} {\rm or} \hspace{0.15cm}  \mu \,. 
\end{equation}
In our analysis, there is no excluded region in the $M_{\nu_R} > M_{W_R}$ region since the corresponding cross section in that region is below the experimental limits, as can be read from the color bars in  \autoref{fig:LRSMneuthalf_2}. This is understood from our previous analysis, as the production cross section of $\nu_R \ell$ through $W_R$ bosons dominates the one obtained through the virtual $W_R^\star$  bosons. The top planes of  \autoref{fig:LRSMneuthalf_2} represent the results of the exclusion in the two dimensional $M_{W_R}-M_{\nu_R}$ mass plane based on the scenario where  
$g_L = g_R$, $\tan \beta = 0.01$ and $V_{\rm CKM}^L = V_{\rm CKM}^R$,  whereas middle and bottom planes show the same exclusion for the scenario where  $g_L \neq g _R=0.37$,  $\tan \beta = 0.5$, $V_{\rm CKM}^L = V_{\rm CKM}^R$ and $g_L \neq g _R=0.37$, $\tan \beta = 0.5$, $V_{\rm CKM}^L \neq V_{\rm CKM}^R$, respectively. For the scenario where $g_L \neq g_R$,  $\tan \beta$ = 0.5 and $V_{\rm CKM}^L = V_{\rm CKM}^R$, $W_R$ bosons with masses up to 3.7 (3.7) TeV in the $ee$ channel and up to 3.7 (3.9) TeV in the $\mu\mu$ channel are excluded at 95\% CL, for $M_{\nu_R}$ up to 2.8 (2.9) TeV in the $ee$ channel and 3.1 (3.0) TeV in the $\mu\mu$ channel. The 2D exclusion limits are less stringent in the $ee$ channel, where $W_R$ boson masses  are excluded up to 3.0 TeV for $\nu_R$ masses close to  the $M_{W_R}=M_{\nu_R}$ degeneracy line. On the other hand, we exclude less parameter space in the two dimensional $M_{W_R}-M_{\nu_R}$ mass plane  when  $g_L \neq  g _R=0.37$,  $\tan \beta= 0.5$  and $V_{\rm CKM}^L \neq V_{\rm CKM}^R$. As seen from the bottom planes of \autoref{fig:LRSMneuthalf_2}, $W_R$ bosons with masses up to 3.1 (3.3) TeV are excluded at 95\% CL, for $M_{\nu_R}$ up to 2.1 TeV,  in the $ee$ channel whereas $W_R$ bosons with masses up to 3.3 (3.4) TeV are excluded at 95\% CL, for $M_{\nu_R}$ up to 2.6 (2.5) TeV in the $\mu\mu$ channel. Here again, the 2D exclusion limits are less stringent in the $ee$ channel, where $W_R$ boson masses are excluded up to 2.0 TeV for $M_{\nu_R}$ masses close to the $M_{W_R}=M_{\nu_R}$ degeneracy line.

%%%%%%%%%%%%%%%%%%%%%%%%%%%%%%%%%%%%%%%%%%%%%%%%%%%%%%
%\subsubsection{ $ 2 M_{\rm \nu_R}  = M_{\rm W_R}$ case}
%\label{subsubsec:LRSMneutrinohalfofWR}
%%%%%%%%%%%%%%%%%%%%%%%%%%%%%%%%%%%%%%%%%%%%%%%%%%%%%
\begin{figure}
	$\begin{array}{cc}
	\includegraphics[scale=0.35]{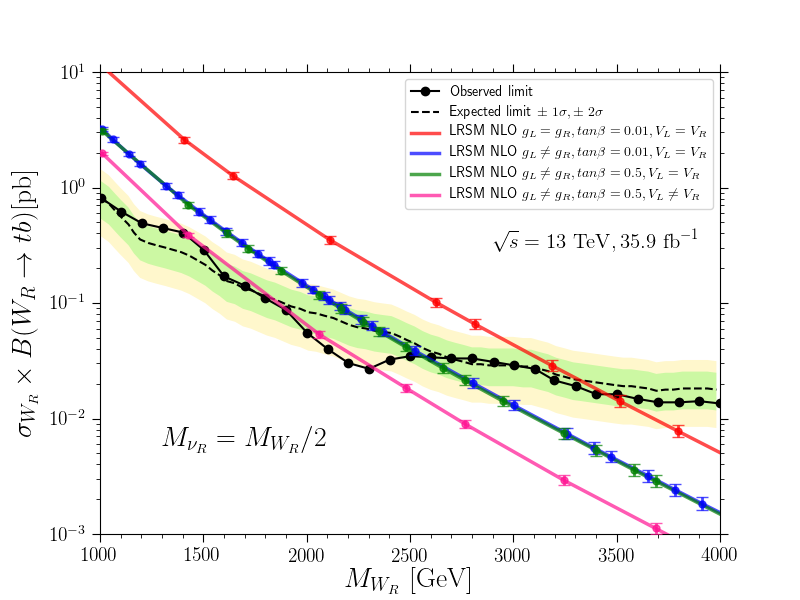}&
	\includegraphics[scale=0.35]{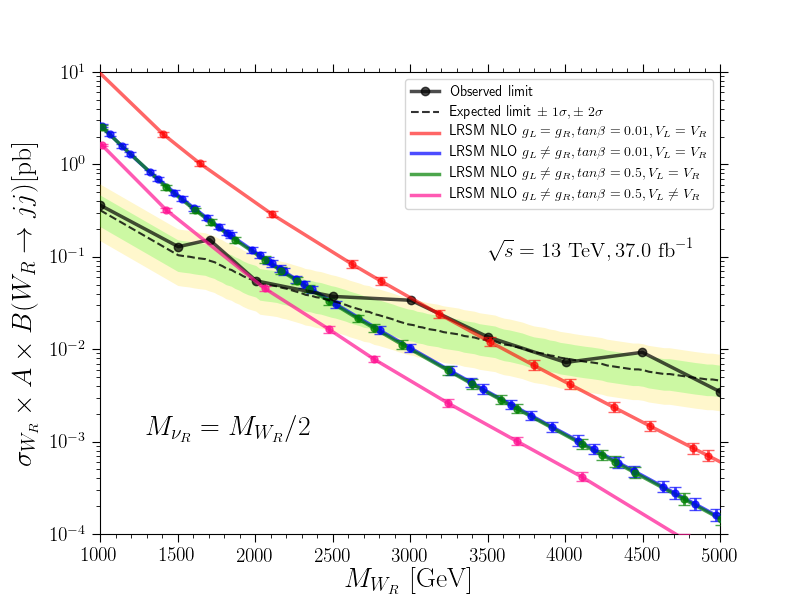}\\
	\includegraphics[scale=0.35]{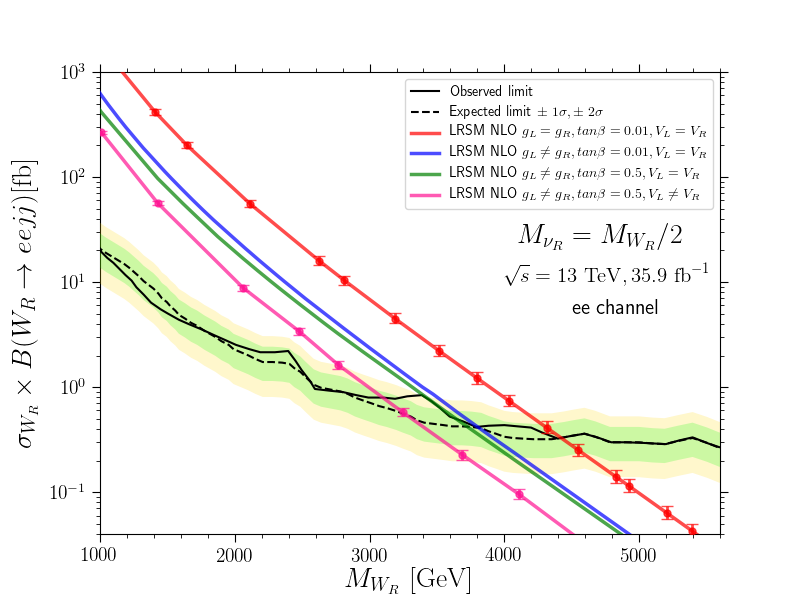}&
	\includegraphics[scale=0.35]{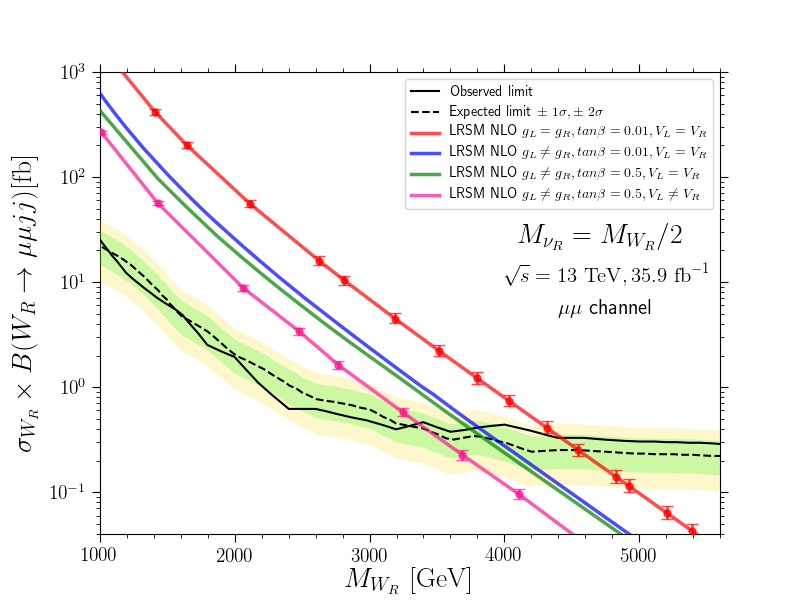}
	\end{array}$
	\caption{(Top left): The cross section of $W_R \to t{\bar b}$ vs $W_R$ mass, compared to CMS data \cite{Sirunyan:2017ukk}; (top right) The cross section of $W_R \to jj$ vs $W_R$ mass, compared to the ATLAS measurements \cite{Aaboud:2018spl}; (bottom left) The cross section of $W_R \to ee jj$ vs $W_R$ mass, for the case  $M_{\nu_R}=M_{W_R}/2$, compared to the CMS data \cite{Sirunyan:2018pom}; (bottom right) The cross section of $W_R \to \mu\mu jj$ vs $W_R$ mass, for the case where $M_{\nu_R}=M_{W_R}/2$, compared to \cite{Sirunyan:2018pom}. }
	\label{fig:LRSMneuthalf}
\end{figure}
%%%%%%%%%%%%%%%%%%%%%%%%%%%%%%%%%%%%%%%%%%%%%%%%%%%%%

\begin{figure}
	$\begin{array}{cc}
	\includegraphics[scale=0.35]{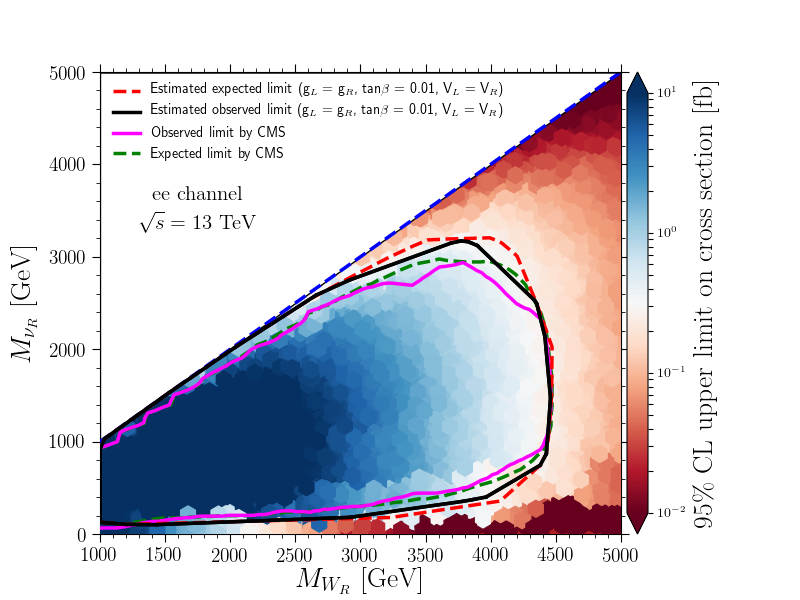} &
	\includegraphics[scale=0.35]{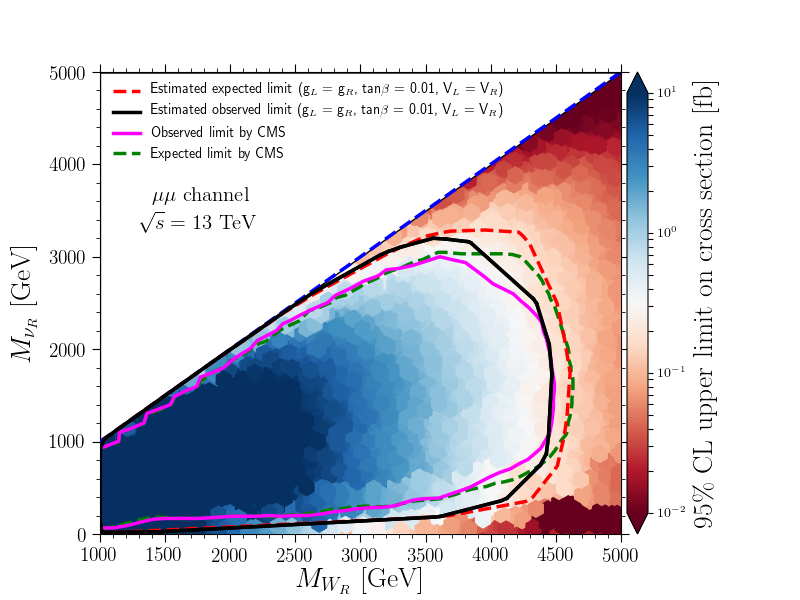}\\
	\includegraphics[scale=0.35]{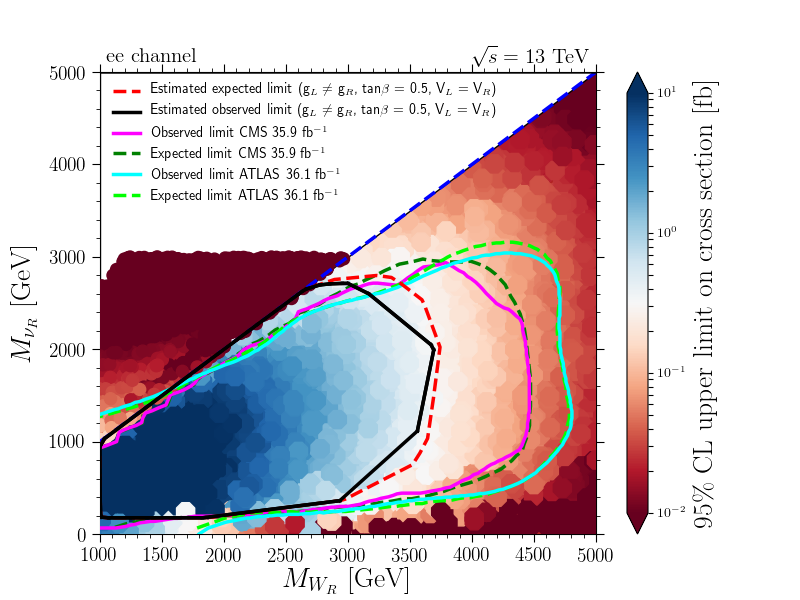}&
	\includegraphics[scale=0.35]{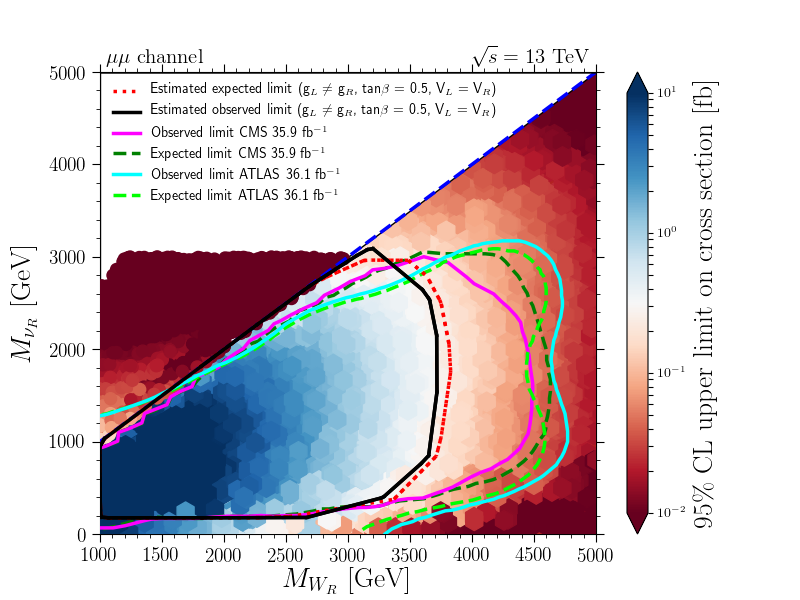}\\
	\includegraphics[scale=0.35]{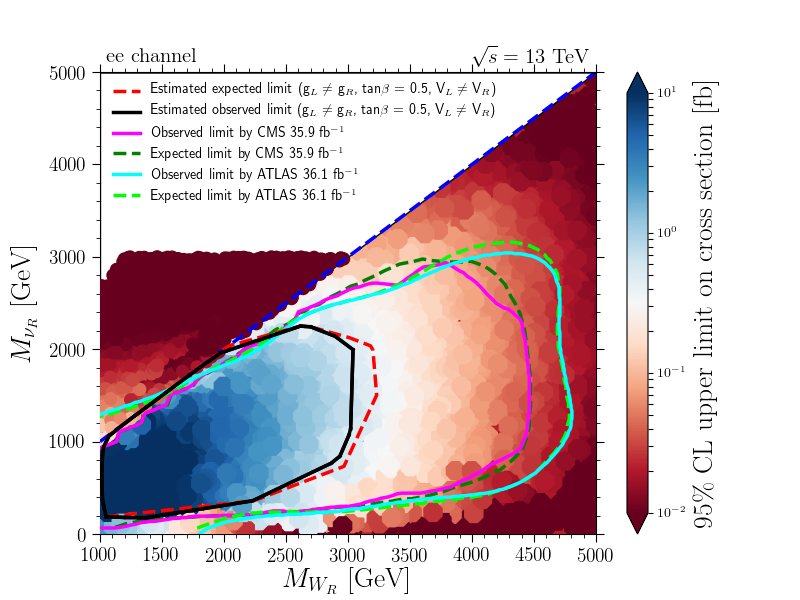}&	
	\includegraphics[scale=0.35]{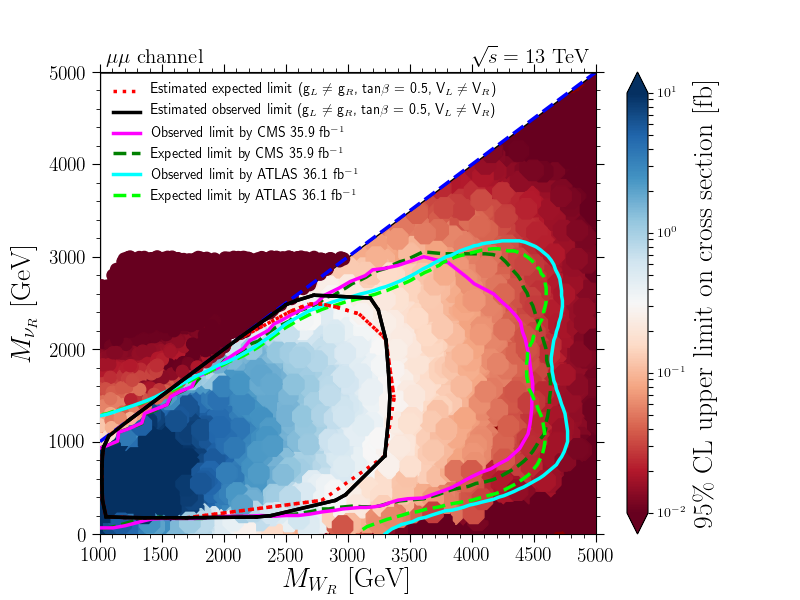}
	\end{array}$
	\caption{Observed (continuous lines) and expected (dashed lines) 95\% CL exclusion contours in the $M_{W_R}-M_{\nu_R}$ plane in the $ee$ (left columns) and $\mu\mu$ (right columns) channels for Majorana $\nu_R$ neutrinos. The dashed blue line in the each plane indicates the region where $M_{W_R} =M_{\nu_R}$. We also show observed and expected limits by ATLAS  \cite{Aaboud:2018spl} @ ${\cal L}=36.1$ fb$^{-1}$, and CMS \cite{Sirunyan:2018pom} @ ${\cal L}=35.9$ fb$^{-1}$,  obtained assuming $g_L =g_R$,  $\tan \beta = 0.01$ and $V_{\rm CKM}^L = V_{\rm CKM}^R$. The cross section values are indicated by the  colors  in the legend of the planes. In the top planes, the continuous (dashed) black (red) line shows the estimated observed (expected) limit for the scenario where  $g_L =g_R$,  $\tan \beta = 0.01$ and $V_{\rm CKM}^L = V_{\rm CKM}^R$. In the middle planes,   $g_L \neq  g_R=0.37$,  $\tan \beta = 0.5$ and $V_{\rm CKM}^L = V_{\rm CKM}^R$,  whereas the same limits in the bottom planes are analyzed for the scenario where    $g_L \neq  g_R=0.37$,  $\tan \beta = 0.5 $,   and also $V_{\rm CKM}^L \neq V_{\rm CKM}^R$.}
	\label{fig:LRSMneuthalf_2}
\end{figure}
In \autoref{tab:ExclusionBoundsI} we show explicitly the limits for different values of $g_L$, $ g_R$,  $\tan \beta $ and $V_{\rm CKM}^L$ versus $  V_{\rm CKM}^R$ values for $M_{\nu_R} > M_{W_R}$ case. The second column gives the expected limits, while the third the observed limits\footnote{Our theoretically estimated observed and expected values are obtained for the points where our model curves intersect the respective observed and expected experimental curves.}. The last column indicates the dominant constraining channel. For this scenario, the most stringent limits come from $W_R \to jj $, for the case when $g_L =  g_R$,  $\tan \beta = 0.01$ and $V_{\rm CKM}^L = V_{\rm CKM}^R$, where $M_{W_R} \ge$ 3.60 (3.625) TeV for the observed (expected) mass limits, while the least constraining case is the decay $W_R \to t {\bar b} $,  when $g_L \neq  g_R=0.37$,  $\tan \beta = 0.5$ and $V_{\rm CKM}^L \neq V_{\rm CKM}^R$, where $M_{W_R} \ge$ 2.36 (1.94) TeV for the observed (expected) mass limits.

We repeat the analysis in \autoref{tab:ExclusionBoundsII} for the case where $M_{\nu_R} < M_{W_R}$, using the same column notations as in \autoref{tab:ExclusionBoundsI}.  For this second scenario, the most stringent limits come from $W_R \to  \mu \mu jj $, for the case when $g_L =  g_R$,  $\tan \beta = 0.01$ and $V_{\rm CKM}^L = V_{\rm CKM}^R$, where $M_{W_R} \ge$ 4.42 (4.5) TeV for the observed (expected) mass limits, while the least constraining case is the case $W_R \to t {\bar b} $,  when $g_L \neq  g_R=0.37$,  $\tan \beta = 0.5$ and $V_{\rm CKM}^L \neq V_{\rm CKM}^R$, where $M_{W_R} \ge$ 2.32 (1.85) TeV for the observed (expected) mass limits.

%\newpage

\begin{table}[]
	\centering
%	\begin{center}
		\small
		\begin{tabular}{P{4.0cm}|P{2.7cm}|P{2.7cm}|P{2.2cm}}
			\hline	\hline
				&\multicolumn{2}{c|}{ }&\\
			%Scenario I: $M_{\nu_R} > M_{W_R} $	& \parbox[t][][t]{3.2cm}{ Expected limits \\ (Estimated)}  & \parbox[t][][t]{3cm}{ Observed limits \\ (Estimated)}  & \parbox[t][][t]{2.5cm}{ Exclusion \\ channel} \\ \hline \hline
	Scenario I: $M_{\nu_R} > M_{W_R} $	& \multicolumn{2}{c|}{Lower limits for $ M_{W_R}$ (GeV)}&Exclusion \\[1mm]
	\cline{2-3}
			&&&channel \\
	&Expected&Observed&\\ [1mm]
	\hline \hline		
			$ g_L = g_R $, $\tan\beta$ = 0.01, $V_{\rm CKM}^L= V_{\rm CKM}^R$	& 3450  & 3600  & $W_R \to tb$  \\ \hline
			$ g_L \neq g_R $, $\tan\beta$ = 0.01, $V_{\rm CKM}^L= V_{\rm CKM}^R$	& 2700  & 2700  & $W_R \to tb$ \\ \hline
			$ g_L \neq g_R $, $\tan\beta$ = 0.5, $V_{\rm CKM}^L= V_{\rm CKM}^R$	& 2675  & 2675  & $W_R \to tb$ \\ \hline
			$ g_L \neq g_R $, $\tan\beta$ = 0.5, $V_{\rm CKM}^L \neq V_{\rm CKM}^R$& 1940 & 2360& $W_R \to tb$ \\ \hline \hline
			
			$ g_L = g_R $, $\tan\beta$ = 0.01, $V_{\rm CKM}^L= V_{\rm CKM}^R$	& 3625  & 3620 & $W_R \to jj$  \\ \hline
			$ g_L \neq g_R $, $\tan\beta$ = 0.01, $V_{\rm CKM}^L= V_{\rm CKM}^R$ & 2700 & 2555  & $W_R \to jj$ \\ \hline
			$ g_L \neq g_R $, $\tan\beta$ = 0.5, $V_{\rm CKM}^L= V_{\rm CKM}^R$	& 2650 & 2500 & $W_R \to jj$ \\ \hline
			$ g_L \neq g_R $, $\tan\beta$ = 0.5, $V_{\rm CKM}^L \neq V_{\rm CKM}^R$	& 2010 & 2000 & $W_R \to jj$ \\ \hline \hline				
		\end{tabular}
	\caption{Lower limits for $M_{W_R}$ in GeV, when $M_{\nu_R} > M_{W_R}$.}
	\label{tab:ExclusionBoundsI}
%	\end{center}
\end{table}
%%%%%%%%%%%%%%%%%%%%%%%%%%%%%%%%%%%%%%%%%%
\begin{table}[]
	\centering
%	\begin{center}
		\small
		\begin{tabular}{P{4.0cm}|P{2.7cm}|P{2.7cm}|P{2.2cm}}
			\hline	\hline
				&\multicolumn{2}{c|}{ }&\\
			%Scenario I: $M_{\nu_R} > M_{W_R} $	& \parbox[t][][t]{3.2cm}{ Expected limits \\ (Estimated)}  & \parbox[t][][t]{3cm}{ Observed limits \\ (Estimated)}  & \parbox[t][][t]{2.5cm}{ Exclusion \\ channel} \\ \hline \hline
	Scenario II: $M_{\nu_R} < M_{W_R} $	& \multicolumn{2}{c|}{Lower  limits for $M_{W_R}$ (GeV)}&Exclusion \\[1mm]
	\cline{2-3}
			&&&channel \\
	&Expected&Observed&\\ [1mm] \hline \hline
			$ g_L = g_R $, $\tan\beta$ = 0.01, $V_{\rm CKM}^L= V_{\rm CKM}^R$	& 4420   & 4420  & $W_R \to qqee $  \\ \hline
			$ g_L \neq g_R $, $\tan\beta$ = 0.01, $V_{\rm CKM}^L= V_{\rm CKM}^R$	& 3800  & 3800  & $W_R \to qqee $ \\ \hline
			$ g_L \neq g_R $, $\tan\beta$ = 0.5, $V_{\rm CKM}^L= V_{\rm CKM}^R$	& 3720 & 3725  & $W_R \to qqee $ \\ \hline
			$ g_L \neq g_R $, $\tan\beta$ = 0.5, $V_{\rm CKM}^L \neq V_{\rm CKM}^R$	& 3300  & 3100  & $W_R \to qqee $ \\ \hline \hline	

			$ g_L = g_R $, $\tan\beta$ = 0.01, $V_{\rm CKM}^L= V_{\rm CKM}^R$	& 4500  & 4420  & $W_R \to qq \mu\mu $  \\ \hline
			$ g_L \neq g_R $, $\tan\beta$ = 0.01, $V_{\rm CKM}^L= V_{\rm CKM}^R$ & 3950 & 3800  & $W_R \to qq\mu \mu $ \\ \hline
			$ g_L \neq g_R $, $\tan\beta$ = 0.5, $V_{\rm CKM}^L= V_{\rm CKM}^R$	& 3900 & 3750  & $W_R \to qq \mu \mu $ \\ \hline
			$ g_L \neq g_R $, $\tan\beta$ = 0.5, $V_{\rm CKM}^L\neq V_{\rm CKM}^R$	& 3400  & 3350  & $W_R \to qq\mu \mu $ \\ \hline \hline				

			$ g_L = g_R $, $\tan\beta$ = 0.01, $V_{\rm CKM}^L= V_{\rm CKM}^R$	& 3320  & 3450 & $W_R \to tb $  \\ \hline
			$ g_L \neq g_R $, $\tan\beta$ = 0.01, $V_{\rm CKM}^L= V_{\rm CKM}^R$	& 2375 & 2575 & $W_R \to tb $ \\ \hline
			$ g_L \neq g_R $, $\tan\beta$ = 0.5, $V_{\rm CKM}^L= V_{\rm CKM}^R$	& 2350 & 2565   & W$_R \to tb $ \\ \hline
			$ g_L \neq g_R $, $\tan\beta$ = 0.5, $V_{\rm CKM}^L\neq V_{\rm CKM}^R$& 1850 & 2320& W$_R \to tb $ \\ \hline \hline				

			$ g_L = g_R $, $\tan\beta$ = 0.01, $V_{\rm CKM}^L= V_{\rm CKM}^R$	& 3500& 3500& $W_R \to jj $  \\ \hline
			$ g_L \neq g_R $, $\tan \beta$ = 0.01, $V_{\rm CKM}^L= V_{\rm CKM}^R$& 2500& 2430& $W_R \to jj $ \\ \hline
			$ g_L \neq g_R $, $\tan\beta$ = 0.5, $V_{\rm CKM}^L= V_{\rm CKM}^R$	& 2460 & 2400& $W_R \to jj $ \\ \hline
			$ g_L \neq g_R $, $\tan\beta$ = 0.5, $V_{\rm CKM}^L\neq V_{\rm CKM}^R$& 2000& 2000& $W_R \to jj $ \\ \hline		
	\end{tabular}
	\caption{Lower limits for $M_{W_R}$ in GeV when $M_{\nu_R} < M_{W_R} $.}
	\label{tab:ExclusionBoundsII}
	%	\end{center}
\end{table}
%
%\clearpage
%%%%%%%%%%%%%%%%%%%%%%%%%%%%%%%%%%%%%%%%%%%%%%%%%%
To assess properties and differences between the two scenarios, in \autoref{tab:freeparamBenchmarks} we give the complete set of parameters for two representative benchmarks, one for the first scenario, $M_{\nu_R} > M_{W_R} $   and one for the second scenario,  $ M_{\nu_R} < M_{W_R} $, with the $W_R$ masses, cross sections and branching ratios given in \autoref{tab:xSectionBenchmark}, for LHC at $\sqrt{s}=13 $ and 27 TeV.

While there are no discerning differences in the parameters of scalar potential, \autoref{eq:pot_htm}, or $\tan \beta$ (chosen to maximize $W_R \to W_L h$), the Yukawa coupling generating the Majorana masses is an order of magnitude larger in Scenario I, needed to generate $ M_{\nu_R} > M_{W_R} $; while in Scenario II $v_R$ is about 50\% larger, to generate an $M_{W_R}$ larger by about 50\% than in Scenario I. The main contributions to the cross sections come from the off-diagonal element $(12)=(us)$ and $(21)=cd$ in $V_{\rm CKM}^R$. For {\bf BM I}, the largest partonic contribution to the cross section ($u {\bar s} \to W_R$)  decreases by 8.2\% from 13 TeV to 27 TeV while the contribution from the same channel decreases by 10.4\% for {\bf BM II}.   For {\bf BM I}, the second largest partonic contribution,  $d {\bar c} \to W_R$ increases by 3.1\% from 13 TeV to 27 TeV, while the contribution from the same channel increases by 5.6\% for {\bf BM II}. Because of the highly non-diagonal  $V_{\rm CKM}^R$ for both benchmarks, the dominant decay is into $jj= q {\bar q}^\prime \neq t {\bar b}$, followed by $ t {\bar b}$ in {\bf BMI}, while in {\bf BM II} the decay into leptons and neutrinos (combined for three families) is of the same strength as $ t {\bar b}$.
%%%%%%%%%%%%%%%%%%%%%%%%%%%%%%%%%%%%%%%%%%%%%%%%%%%%
\begin{table}[]
	\begin{center}
		%\large
		\begin{tabular}{|c|c|c|c|c|c|c|c|c|}
			\hline
		& $v_R$ & $g_R$ & $\tan\beta$ & diag(h$^{ij}_R$) & $\alpha_1$ & $\alpha_2$ & $\alpha_3$    \\ \hline
  $\bold{BM\ I :}$	& 9.8 TeV  & 0.36 & 0.55  & 0.9   & 0.6  & 0.6  & 2.0   \\ \hline
  $\bold{BM\ II :}$	& 14.1 TeV & 0.37  & 0.55 & 7.5 $\times 10^{-2}$  & 0.6 & 0.6 & 2.0    \\ \hline \hline
		& $\lambda_1$  & $\lambda_2$  & $\lambda_3$ & $\rho_1$  & $\rho_2$  & $\rho_3$ & $\rho_4=\lambda_4$  \\ \hline
   $\bold{BM\ I :}$	& 0.15  & 0.14 & 0.142 & 2.12$\times 10^{-3}$ & 3.4$\times 10^{-3}$  & 5.5$\times 10^{-3}$ & 0.0   \\ \hline
   $\bold{BM\ II :}$ & 0.16 & 0.16 & 0.162 & 1.79$\times 10^{-3}$ & 4.0$\times 10^{-3}$  & 4.0$\times 10^{-3}$  & 0.0 \\ \hline
		\end{tabular}
	\caption{Parameter values for {\bf BM I} and {\bf BM II}. }
	\label{tab:freeparamBenchmarks}
	\end{center}
\end{table}

\begin{table}[]
	\begin{center}
		%\large
		\begin{tabular}{c||c|c}
			\hline
			& $\bold{BM\ I :}~ M_{\nu_R} > M_{W_R} $   &   $\bold{BM\ II :}~ M_{\nu_R} < M_{W_R} $   \\ \hline \hline
			$m_{W_R}$ [GeV]                        & 2557  & 3689 \\ \hline
			$m_{\nu_R}$ [GeV]                                 & 16797  & 1838 \\ \hline
			$\sigma$(pp $\to$ $W_R $)  [fb] @13 TeV & 48.7  & 3.98  \\ \hline
			$\sigma$(pp $\to$ $W_R $)  [fb] @27 TeV & 478.0  & 77.3  \\ \hline \hline
			BR($W_R \to  t\overline{b}$) [\%] & 26.3 & 19.9 \\ \hline 
			BR($W_R \to  jj$) [\%] & 58.6 & 45.8 \\ \hline 
			BR($W_R \to  \nu_R \ell $) [\%] & - & 6.5 (each family) \\ \hline 	
			BR($W_R \to  h_1 W_L $) [\%] & 1.8 & 1.5 \\ \hline 
	    	BR($W_R \to  W_L Z $) [\%] & 2.0 & 1.6  \\ \hline \hline
			BR($\nu_R \to  \ell qq^\prime $) [\%] & - & 65.3 \\ \hline 		
			BR($\nu_R \to  W_ L \ell$) [\%] & 1.1$\times 10^{-4}$ & 33.1 \\ \hline 
			BR($\nu_R \to  W_ R \ell$) [\%] & 99.9 & - \\ \hline \hline
%			$\epsilon_K$ &   & 2.84 $\times 10^{-3}$ \\ \hline 
%			$\epsilon_K/\epsilon_K^{SM}$ &   & 0.86 \\ \hline
%			$\Delta M_K$ &   & 2.46 $\times 10^{-15}$ \\ \hline 			 
%			$\Delta M_K/\Delta M_K^{SM}$ &   & 1.16 \\ \hline 
%			$\Delta B_s$ &   & 12.5 \\ \hline 
%			$\Delta B_d$ &  &  0.68 \\ \hline
%			BR($B \to  X_s \gamma$) &  & 4.13 $\times 10^{-4}$ \\ \hline \hline			 					%															
		\end{tabular}
		\caption{Related Branching Ratios and Cross Sections for {\bf BM I} and {\bf BM II}.}
		\label{tab:xSectionBenchmark}
	\end{center}
\end{table}
%\clearpage
%\newpage
%%%%%%%%%%%%%%%%%%%%%%%%%%%%%%%%%%%%%%%%%%%%%%%%%%%%%
\section{Summary and Conclusion}
\label{sec:conclusion}
We performed a comprehensive analysis of mass bounds on $W_R$ gauge boson in a  LRSM with general gauge couplings, VEVs, mixing angles and right-handed quark mixing matrix. We reinterpret the CMS and ATLAS analyses using this scenario, and find considerably relaxed bounds on $W_R$ masses. 
We divide the parameter space into two cases, one for $M_{W_R}< M_{\nu_R}$, and another for $M_{W_R}>M_{\nu_R}$, as this ordering has implications on $W_R$ decay channels.

When $M_{\nu_R} >M_{W_R}$, decays of right-handed $W$-boson  into a top quark and a bottom quark or  into  two jets are dominant. We compare our predictions with the results of CMS and ATLAS at $\sqrt{s}$ = 13 TeV, integrated luminosity of 35.9 fb$^{-1}$ and 37.0 fb$^{-1}$, respectively. 95\% CL upper limits on the product of the $W_R$ boson production cross section with its branching fraction to a top and a bottom quark ($t \bar b$) or  with the branching fraction to two jets ($jj$) are calculated as a function of the $W_R$ boson mass.  These analyses recover the excluded ranges obtained by CMS and ATLAS ranging from 55.4\% for the two jets channel to 65.5\% for a top and bottom quark channel in the limit of exact left-right symmetry and small $\tan \beta$, while when relaxing these limits, much less  stringent constraints are obtained on the production of $W_R$ bosons decaying  to a top and a bottom quark and on $W_R$ bosons decaying into two jets. Note that increasing $\tan \beta$ only serves to enhance the non-fermionic decays $W_R \to W_L h$ and $W_R \to W_L Z_L$, whose branching ratios are small for allowed values of $\tan \beta$, 2\% at most. While the results are not particularly sensitive to $\tan \beta$, they depend crucially on $g_R$ and $V_{\rm CKM}^R$. The least constrained choice for this scenario is  setting  $g_L \neq  g_R=0.37$,  $\tan \beta = 0.5 $ and $V_{\rm CKM}^L \neq   V_{\rm CKM}^R$, when the observed (expected) lower limit on $W_R$ mass at 95\% CL  is 2360 (1940) GeV for  the $tb$ channel and 2000 (2010) GeV for the $jj$ channel.

When $M_{\nu_R} <M_{W_R}$, we investigate  exclusion bounds for $W_R$ bosons and heavy right-handed Majorana neutrinos masses, using the final states containing a pair of charged leptons (electrons or muons), and two jets ($\ell\ell jj$), with $\ell = e, \mu$. We compare our result with the results obtained by CMS (ATLAS) with  ${\cal L}=35.9$ fb$^{-1}$ (36.1 fb$^{-1}$) at $\sqrt{s} = 13$ TeV at the LHC. We again reproduce the CMS and ATLAS results for $g_L =  g_R$,  $\tan \beta = 0.01 $ and $V_{\rm CKM}^L =  V_{\rm CKM}^R$, and find  three different parameter choices which relax the current $W_R$ and $\nu_R$ mass exclusion limits. Assuming that gauge couplings and CKM matrices in the right-handed sector are different from those of the SM, a region in the two-dimensional plane ($M_{W_R}-M_{\nu_R}$) excluding a smaller range of right-handed neutrino masses is found at 95\% CL. Based on the scenario where we allow $g_L \neq  g_R=0.37$,  $\tan \beta = 0.5 $ and $V_{\rm CKM}^L \neq   V_{\rm CKM}^R$, a $W_R$ boson decaying into a right-handed heavy neutrino  is excluded at 95\% CL up to a mass of 3100 GeV in the $ee$ channel,  and 3350 GeV in the $\mu\mu$ channel, providing less stringent limits on this parameter space. The excluded region for $M_{W_R}$ extends to about  $M_{W_R} \simeq$ 3.1 TeV, for $M_{\nu_R}$ $\simeq$ 2.0 TeV in the electron channel and about $M_{W_R} \simeq$ 3.3 TeV, for $M_{\nu_R}$ $\simeq$ 2.0 TeV in the muon channel. Conversely, the $M_{\nu_R}$ limits reach about 2.2 TeV for $M_{W_R}$ = 2.6 TeV in the electron channel and 2.6 TeV for $M_{W_R}$ = 2.8 TeV in the muon channel. We also analyze the region where $M_{\nu_R} >M_{W_R}$ in the $M_{W_R}-M_{\nu_R}$ contour plot, but find no excluded solutions in this region. These results also  recover the experimentally excluded ranges, further relaxing the limits obtained by CMS and ATLAS, ranging from 66\% for BR's in the $eejj$ channel to 71\% for $\mu\mu jj $ channel.  Overall our results yield weaker limits on $W_R$ mass from the production of $W_R$ bosons decaying into $\ell \ell jj$, yielding hope that $W_R$ could be discovered at HL-HE LHC.

Finally, allowing $g_L \neq  g_R=0.37$  and $V_{\rm CKM}^L \neq   V_{\rm CKM}^R$ will not have significant consequences on other sectors of the model. Both the singly charged and doubly charged Higgs bosons $\delta_R^+$ and $\delta_R^{++}$ are expected to be heavy. Even so, their production mechanism  is dominated by photon-mediated Drell Yan, or $\gamma \gamma$ fusion, and their branching ratios are independent of $g_R$. This leads further support to $W_R$ production and  decay as being most promising signal to test this scenario.
%%%%%%%%%%%%%%%%%%%%%%%%%%%%%%%%%%%%%%%%%%%%%%%%%%%%%
\begin{acknowledgments}
%%%%%%%%%%%%%%%%%%%%%%%%%%%%%%%%%%%%%%%%%%%%%%%%%%%%%
  Part of numerical calculations reported in this paper was performed using High Performance Computing (HPC), managed by Calcul Qu{\'e}bec and Compute Canada. We gratefully acknowledge Shastri Institute for a Canada-India collaboration grant (SRG 2017-18). M.F. and \"{O}.\"{O}. also thank NSERC for partial financial support under grant number SAP105354.
\end{acknowledgments}
%%%%%%%%%%%%%%%%%%%%%%%%%%%%%%%%%%%%%%%%%%%%%%%%%%%%%%%%%%%%%
\bibliography{WRexcl}

\end{document}